\documentclass[pra,twocolumn,superscriptaddress,showpacs,aps]{revtex4}

\usepackage{amsfonts}
\usepackage{amsmath}
\usepackage{amssymb}
\usepackage{graphicx}
\usepackage{dcolumn}
\usepackage{times}
\usepackage[toc,page]{appendix}

\DeclareMathOperator{\sech}{sech}

\begin{document}

\title{Multiple dark-bright solitons in atomic Bose-Einstein condensates}

\author{D.~Yan}
\affiliation{Department of Mathematics and Statistics, 
University of Massachusetts,
Amherst, Massachusetts 01003-4515, USA}
\author{J.J.~Chang}
\affiliation{Washington State University,
Department of Physics \& Astronomy,
Pullman, Washington 99164, USA}
\author{C.~Hamner}
\affiliation{Washington State University,
Department of Physics \& Astronomy,
Pullman, Washington 99164, USA}
\author{P.G.~Kevrekidis}
\affiliation{Department of Mathematics and Statistics, University of Massachusetts,
Amherst, Massachusetts 01003-4515, USA}
\author{P.~Engels}
\affiliation{Washington State University,
Department of Physics \& Astronomy,
Pullman, Washington 99164, USA}
\author{V.~Achilleos}
\affiliation{Department of Physics, University of Athens, Panepistimiopolis,
Zografos, Athens 157 84, Greece}
\author{D.J.~Frantzeskakis}
\affiliation{Department of Physics, University of Athens, Panepistimiopolis,
Zografos, Athens 157 84, Greece}
\author{R.~Carretero-Gonz\'alez}
\affiliation{Nonlinear Dynamical Systems
Group\footnote{\texttt{URL}: http://nlds.sdsu.edu},
Department of Mathematics and Statistics, 
and Computational Science Research Center,
San Diego State University, San Diego,
California 92182-7720, USA}
\author{P.~Schmelcher}
\affiliation{ Zentrum f\"ur Optische Quantentechnologien, Universit\"at
Hamburg, Luruper Chaussee 149, 22761 Hamburg, Germany}

\begin{abstract}


We present experimental results and a systematic theoretical analysis of dark-bright soliton interactions and multiple-dark-bright soliton complexes in atomic two-component Bose-Einstein condensates. We study analytically the interactions between two-dark-bright solitons in a homogeneous condensate and, then, extend our considerations to the presence of the trap.
An effective equation of motion is derived for the dark-bright soliton center
and the existence and stability of stationary two-dark-bright soliton states
is illustrated (with the bright components being either in- or out-of-phase).
The equation of motion provides the characteristic oscillation frequencies of
the solitons, in good agreement with the eigenfrequencies of the anomalous
modes of the system.

\end{abstract}

\maketitle

\section{Introduction}

Over the past few years, the macroscopic nonlinear structures that can be supported in atomic Bose-Einstein condensates (BECs) have been a topic of intense investigation (see, e.g., Refs.~\cite{emergent,revnonlin,rab,djf} for reviews in this topic). The first experimental efforts to identify the predominant nonlinear structure in BECs with repulsive interatomic interactions, namely the dark soliton, were initiated over a decade ago \cite{han1,nist,dutton,bpa,han2}. However, these efforts suffered from a number of instabilities arising due to dimensionality and/or temperature effects. More recently, a new generation of relevant experiments has emerged, that has enabled the overcoming (or quantification) of some of the above limitations. The latter works have finally enabled the realization of oscillating, and even interacting, robust dark solitons in atomic BECs. This has been achieved by means of various techniques, including phase-imprinting/density engineering \cite{hamburg,hambcol,technion}, matter-wave interference \cite{kip,kip2}, or dragging localized defects through the BECs \cite{engels}.

Atomic dark solitons may also exist in multi-component condensates, where they are coupled with other nonlinear macroscopic structures \cite{emergent,revnonlin,djf}. Of particular interest are dark-bright (DB) solitons that are supported in two-component \cite{BA} and spinor \cite{DDB} condensates. Such structures, are frequently called ``symbiotic'' solitons, as the bright-soliton component (which is generically supported in BECs with attractive interactions \cite{rab}) may only exist due to the inter-species interaction with the dark-soliton component. Dark-bright solitons have also attracted much attention in other contexts, such as nonlinear optics~\cite{yuri} and mathematical physics~\cite{ablowitz}. In fact, DB-soliton states were first observed in optics experiments, where they were created in photorefractive crystals \cite{seg1}, while their interactions were partially monitored in Ref.~\cite{seg2}. In the physics of BECs, robust DB-solitons were first observed in the experiment of Ref.~\cite{hamburg} by means of a phase-imprinting method, and more recently in Refs.~\cite{engels1,engels2,engels3} by means of the counterflow of the two BEC components. The above efforts led to a renewed interest in theoretical aspects of this theme: this way, DB-soliton interactions were studied from the viewpoint of the integrable systems theory in Ref.~\cite{rajendran}, DB-soliton dynamics were investigated numerically in Ref.~\cite{berloff}, while DB-solitons in discrete settings were recently analyzed in Ref.~\cite{azucena}. Furthermore, higher-dimensional generalizations ---namely, vortex-bright-soliton structures--- were recently studied as well~\cite{VB}.

Our aim in the present work is to study multiple-DB solitons in two-component BECs confined in harmonic traps. First, we present our experimental results, based on the counterflow of two rubidium condensate species \cite{engels1,engels2,engels3}, which demonstrate the existence (and indicate the robustness) of such multiple-DB-soliton states. Motivated by the experimental observations, we then proceed to analyze the interactions of DB solitons, first in the case of a homogeneous system and, next, in the presence of the trap. Our analytical approximation relies on a Hamiltonian perturbation theory, which leads to an equation of motion of the centers of  DB-soliton interacting pairs. Employing this equation of motion, we demonstrate the existence of  stationary two- and three-DB-soliton states, find semi-analytically the equilibrium distance of the constituent solitons, as well as the oscillation frequencies around these equilibria. The 
oscillation frequencies correspond to the characteristic anomalous modes' eigenfrequencies that we compute via a Bogoliubov-de Gennes (BdG) analysis. This way, we are able to quantify the properties of stationary multiple-DB-solitons in harmonically confined two-component BECs, and provide analytical results for their in- and out-of-phase motions.

The paper is organized as follows. In Section~II, we provide our experimental results. In Section III we describe our theoretical setup and present the DB-soliton states. Section~IV is devoted to the study of the interactions of two DB-solitons, while Section~V contains the results for multiple DB-solitons in the trap. Finally, in Section~VI we summarize our findings and discuss future challenges.

\section{Experimental results - Motivation}

\begin{figure}
\centering
\includegraphics[width=8.5cm]{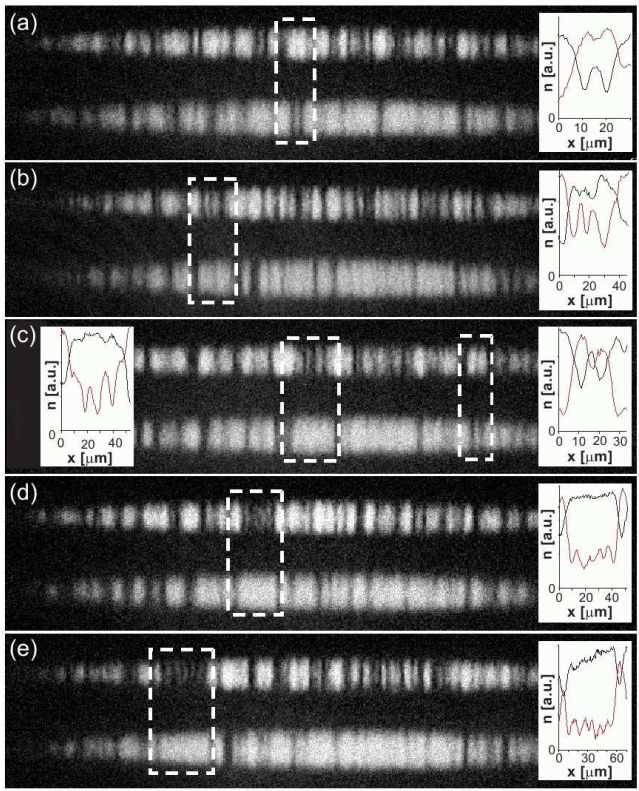}
\caption{(Color online)
Experimental images indicating DB-soliton clustering in a two-component BEC.
The upper cloud in each image (and red curve in inset) shows atoms in
the $|2, -2\rangle$ state, while the lower cloud (black curve) shows atoms
in the $|1, -1\rangle$ state.
Prior to imaging, the two components are overlapped in trap for 5~sec.
Insets show integrated cross sections of the boxed region.
For details see text.}
\label{fig6}
\end{figure}

Since our scope is the study of multiple-DB-solitons in atomic BECs, we start by presenting some experimental results, which showcase the existence of such structures. These results, apart from being interesting in their own right ---as they demonstrate the formation of DB-soliton clusters in two-component BECs---
provide the motivation for a systematic analysis of multiple-DB-solitons,
which will be presented in the following sections.

Our soliton generation scheme is based on a counterflow induced modulational instability details of which have been described in Refs.~\cite{engels1,engels2}. Briefly, we start with a BEC of about $800,000$ $^{87}$Rb atoms in the
$|F, m_F \rangle =|1, -1\rangle$ hyperfine state. The atoms are confined in an elongated optical dipole trap with measured trap frequencies of $2\pi\times$\{1.5, 140, 178\}~Hz. About half the atoms are then transferred to the $|2, -2\rangle$ state with a brief microwave sweep, thus producing a weakly miscible two-component mixture. Subsequently, a magnetic gradient of 10.4~mG/cm is applied along the elongated axis of the BEC, inducing counterflow of the two components. As a result, a dense modulation instability pattern arises. Once the pattern has fully developed, the gradient is turned off. During the subsequent in-trap evolution, the initially very regular pattern becomes irregular, and over a time scale
of several seconds displays the dynamics of interacting solitons
originating from the modulational instability. Images taken in several experimental runs 5~sec after the switch-off of the gradient are presented in Fig.~\ref{fig6}, in which the upper cloud (and the red curves in the insets) in each image shows atoms in the $|2, -2\rangle$ state after 7~ms of free expansion, while the lower cloud (and black curves in the insets) shows the atoms in the $|1, -1\rangle$ state after 8~ms of expansion. This difference in expansion time just serves to separate the two states vertically for imaging.

An intriguing observation is the frequent formation of large gaps in one component (which constitutes the component supporting the dark-solitons) that are filled by bright-solitons in the other component. Interestingly, these gaps are structured by small, periodic density bumps, indicating that these regions are composed of merged solitons. Some of these features are marked by the boxed regions in Fig.~\ref{fig6}, with corresponding cross sections shown as insets. We clearly observe clusters of two- and three-merged solitons [see Fig.~\ref{fig6}(a-c)], and also have some indications of clusters composed of four- to five-solitons ---see Fig.~\ref{fig6}(d, e).

While our destructive imaging technique does not allow us to analyze the dynamics and lifetime of the clusters in detail, the occurrence of large DB-soliton clusters strongly supports the theoretical part of our work that we will present below: in fact, we will study analytically the interaction between two-DB solitons, and we will demonstrate the existence of two- and multiple-DB stationary states, resembling the ones observed in the experiment. Furthermore, we will study the stability of these states and discuss their dynamics in the presence of the harmonic trap.

\section{Model and theoretical setup}

\subsection{Coupled GPEs and dark-bright solitons}

Following the experimental observations of the previous section, we consider a two-component elongated (along the $x$-direction) BEC, composed of two different hyperfine states of rubidium. As is the case of the experiment, we consider a highly anisotropic trap, with the longitudinal and transverse trapping frequencies such that $\omega_x \ll \omega_{\perp}$. In the framework of the mean-field theory, the dynamics of this two-component BEC can be described by the following system of two coupled GPEs \cite{emergent,revnonlin,djf}:
\begin{eqnarray}
\!\!\!i\hbar \partial_t \psi_j
\!\! =\!\!
\left( -\frac{\hbar^2}{2m} \partial_{x}^2 \psi_j +V(x) -\mu_j + \sum_{k=1}^2 g_{jk} |\psi_k|^2\right)\!\psi_j.
\label{model}
\end{eqnarray}
Here, $\psi_j(x,t)$ ($j=1,2$) denote the mean-field wave functions of the two components (normalized to the
numbers of atoms $N_j = \int_{-\infty}^{+\infty} |\psi_j|^2 dx$), $m$ is the atomic mass, $\mu_j$ are the chemical potentials,
and $V(x)$ represents the external harmonic trapping potential, $V(x)=(1/2)m\Omega^2 x^2$ where
$\Omega=\omega_x/\omega_\perp$. In addition, $g_{jk}=2\hbar\omega_{\perp} a_{jk}$ are the effective 1D coupling constants, $a_{jk}$ denote the three
$s$-wave scattering lengths (note that $a_{12}=a_{21}$) accounting for collisions between atoms belonging to the same ($a_{jj}$) or different ($a_{jk}, j \ne k$) species. In the case of the hyperfine states $|1,-1\rangle$ and $|2,-2\rangle$ of $^{87}$Rb considered in the previous section, the scattering lengths take the values $a_{11}=100.4a_0$, $a_{12}=98.98a_0$ and $a_{22}=98.98a_0$ (where $a_0$ is the Bohr radius) \cite{engels1,engels2}. Thus, we may safely use the 
approximation that all scattering lengths take the same value, say $a_{ij} \approx a$~\footnote{I.e., the results that will be presented below do not significantly change in the case of unequal $a_{11}$; however, the analytical calculations and formulas given herein are considerably more involved.}. To this end, measuring the densities $|\psi_j|^2$, length, time and energy in units of $2a$, $a_{\perp} = \sqrt{\hbar/\omega_{\perp}}$, $\omega_{\perp}^{-1}$ and $\hbar\omega_{\perp}$,
respectively, we may reduce the system of Eqs.~(\ref{model}) into the following dimensionless form,
\begin{eqnarray}
i \partial_t \psi_j  =&-&\frac{1}{2} \partial_{x}^2 \psi_j  + V(x) \psi_j \nonumber \\
&+&(|\psi_j|^2 + |\psi_{3-j}|^2 -\mu_j) \psi_j, \quad j=1,2.
\label{deq}
\end{eqnarray}
Below, we will consider a situation where the component characterized by the wavefunction $\psi_1$ ($\psi_2$) supports a single- or a multiple-dark (bright) soliton state, and the respective chemical potentials will be such that $\mu_1>\mu_2$. Note that the external potential in Eqs.~(\ref{deq}) takes the form $V(x)=(1/2)\Omega^{2} x^{2}$, where $\Omega = \omega_x/\omega_\perp \ll 1$ is the normalized trap strength.

We assume that a single- or a multiple-dark-soliton state is on top of a Thomas-Fermi (TF) cloud with density $|\psi_{\rm TF}|^2 = \mu_1 -V(x)$; this way, the density $|\psi_1|^2$ in Eqs.~(\ref{deq}) is substituted as $|\psi_1|^2 \rightarrow |\psi_{\rm TF}|^2 |\psi_1|^2$. Furthermore, we introduce the transformations $t \rightarrow \mu_1 t$, $x \rightarrow {\sqrt{\mu_1}}x$,
$|\psi_2|^2 \rightarrow \mu_{1}^{-1} |\psi_2|^2$, and cast Eqs.~(\ref{deq}) into the following form:
\begin{eqnarray}
&&i \partial_{t}\psi_1  +\frac{1}{2} \partial_{x}^2 \psi_1  -(|\psi_1|^2 +  |\psi_2|^2 -1) \psi_1 = R_1,
\label{eq1d} \\
&&i \partial_{t} \psi_2  +\frac{1}{2} \partial_{x}^2\psi_2 - (|\psi_1|^2 + |\psi_2|^2-\tilde{\mu}) \psi_2 = R_2,
\label{eq2d}
\end{eqnarray}
where $\tilde{\mu} = \mu_2/\mu_1$, while
\begin{eqnarray}
R_1 &\equiv& (2\mu_1^2)^{-1}\left[2(1-|\psi_1|^2)V(x)\psi_1 +V'(x)\partial_x \psi_1 \right],
\nonumber
\\[1.0ex]
R_2 &\equiv &\mu_1^{-2}\left[(1-|\psi_1|^2)V(x)\psi_2 \right],
\label{RdRb}
\end{eqnarray}
with $V'(x)\equiv dV/dx$. Equations (\ref{eq1d})-(\ref{eq2d}) can be viewed as a system of two coupled perturbed nonlinear Schr\"{o}dinger (NLS) equations, with perturbations given by Eqs.~(\ref{RdRb}). In the absence of the trap (i.e., for $\Omega=0$), the perturbations vanish and Eqs.~(\ref{eq1d})-(\ref{eq2d}) actually constitute the completely integrable Manakov system \cite{Manakov}. This system conserves, among other quantities, the Hamiltonian (total energy),
\begin{eqnarray}
E &=& \frac{1}{2}\int_{-\infty}^{+\infty} \mathcal{E} dx, \nonumber \\
\mathcal{E} &=& |\partial_{x} \psi_1|^2+|\partial_{x} \psi_2|^2+(|\psi_1|^2+|\psi_2|^2-1)^2 \nonumber \\
&-&2(\tilde{\mu}-1)|\psi_2|^2,
\label{energy}
\end{eqnarray}
as well as the total number of atoms,
$N=N_1+N_2=\sum_{j=1}^2 \int_{-\infty}^{+\infty} |\psi_j|^2 dx$; additionally, the number of atoms of each component, $N_1$ and $N_2$, is separately conserved.

Considering the boundary conditions $|\psi_1|^2 \rightarrow 1 $ and $|\psi_2|^2 \rightarrow 0$ as $|x| \rightarrow \infty$, the NLS Eqs.~(\ref{eq1d})-(\ref{eq2d}) possess an exact analytical single-DB soliton solution of the following form (see, e.g., Ref.~\cite{BA}):
\begin{eqnarray}
\psi_1(x,t)&=&\cos\phi\tanh\left[D(x-x_0(t)\right]+i\sin\phi,
\label{dsoliton2}
\\[1.0ex]
\psi_2(x,t)&=&\eta\sech\left[D(x-x_0(t)\right]\exp\left[ikx+i\theta(t)\right],
\label{bsoliton2}
\end{eqnarray}
where $\phi$ is the dark soliton's phase angle, $\cos\phi$ and $\eta$ represent the amplitudes of the dark and
bright solitons, $D$ and $x_0(t)$ denote the width and the center of the DB soliton, while
$k=D\tan\phi = {\rm const.}$ and $\theta(t)$ are the wavenumber and phase of the bright soliton,
respectively. The above parameters of the single DB-soliton are connected through the following equations:
\begin{eqnarray}
D^2&=& \cos^2\phi-\eta^2,
\label{width} \\[1.0ex]
\dot{x}_0 &=& D\tan\phi,
\label{x0} \\
\theta(t)&=&\frac{1}{2}(D^2-k^2)t+(\tilde{\mu}-1)t,
\label{omegat}
\end{eqnarray}
where $\dot{x}_0=d x_0/dt$ is the DB soliton velocity. Below, we will mainly focus on stationary solutions, characterized by a dark soliton's phase angle $\phi = 0$ [in this case, the bright soliton component is stationary as well ---see Eq.~(\ref{x0})]; nevertheless, we will also consider the near-equilibrium motion of DB solitons, characterized by $\phi \approx 0$.

To describe a two-DB-soliton state (for $\Omega=0$) composed by a pair of two equal-amplitude single DB solitons traveling in opposite directions, we will use the following ansatz:
\begin{eqnarray}
\psi_1(x,t)
&=&      \left(\cos\phi\tanh X_{-}+i\sin\phi \right) \nonumber \\
&\times& \left(\cos\phi\tanh X_{+}-i\sin\phi\right),
\label{eq15}
\\[1.0ex]
\psi_2(x,t)
&=& \eta\, {\rm sech} X_{-}\, e^{i\left[+kx+\theta(t)+(\tilde{\mu}-1)t\right]} \nonumber \\
&+& \eta\, {\rm sech} X_{+}\, e^{i\left[-kx+\theta(t)+(\tilde{\mu}-1)t\right]}\,
e^{i\Delta\theta},
\label{eq16}
\end{eqnarray}
where $X_{\pm} = D\left(x \pm x_0(t)\right)$, $2x_0$ is the relative distance between the two solitons, and $\Delta\theta$ is the relative phase between the two bright solitons (assumed to be constant); below we will consider both the out-of-phase case, with $\Delta\theta=\pi$, as well as the in-phase case, corresponding to $\Delta\theta=0$.

Notice that in either cases of single- or multiple-DB-solitons, the number of atoms of the bright soliton, $N_2$, may be used to connect the amplitude $\eta$ of the bright soliton(s), the chemical potential $\mu_1$ of the dark soliton(s) component, as well as the width $D$ of the DB soliton. In particular, in the case of a single-DB-soliton, one finds that $N_2= 2\eta^2 \sqrt{\mu_1}/D$ [for the variables appearing in Eqs.~(\ref{deq})], while for the case of a two-DB-soliton state (with well-separated solitons) the relevant result is approximately twice as large, namely:
\begin{equation}
N_2 \approx \frac{4\eta^2 \sqrt{\mu_1}}{D}.
\label{Nbmod}
\end{equation}


\subsection{Stationary states and their excitation spectrum}

 In our numerical computations, we will initially obtain (by means of a fixed-point algorithm) stationary solutions of Eqs.~(\ref{deq}), in the form $\psi_1(x,t)=u(x)$ and $\psi_2(x,t)=v(x)$, and then consider their linear stability. This is numerically studied via the well-known BdG analysis (see, e.g., Refs.~\cite{emergent,revnonlin,djf}), upon introducing the following ansatz into Eqs.~(\ref{deq}),
\begin{eqnarray}
\psi_1(x,t)&=& u(x) + \varepsilon \left[a(x) e^{\lambda t}
+ b^{\ast}(x) e^{\lambda^{\ast} t} \right],
\label{eq6}
\\
\psi_2(x,t)&=& v(x) + \varepsilon \left[c(x) e^{\lambda t}
+ d^{\ast}(x) e^{\lambda^{\ast} t} \right].
\label{eq7}
\end{eqnarray}
The resulting equations are linearized (keeping only terms of order of the
small parameter $\varepsilon$), and the  ensuing eigenvalue problem for
eigenmodes $\{a(x), b(x), c(x), d(x)\}$ and eigenvalues
$\lambda= \lambda_r + i \lambda_i$ is solved [note that the asterisk in Eqs.~(\ref{eq6})-(\ref{eq7}) denotes complex conjugation]. In the case of a single DB soliton, the excitation spectrum can be well-understood in both cases, corresponding to the absence and the presence of the harmonic trap, using the following
arguments.

First, in the absence of the trap, the system of Eqs.~(\ref{deq}) features not only a $U(1)$ (phase) invariance in each of the components but also a translational invariance; thus, the system has three pairs of eigenvalues (each associated with one of the above symmetries) at the origin of the spectral plane $(\lambda_r,\lambda_i)$. In this case, the phonon band (associated with the continuous spectrum of the problem) covers the entire imaginary axis of the spectral plane.

Second, in the presence of the trap, the single DB soliton ``lives'' on the background of the confined ground state, i.e., $\{\psi_1, \psi_2\}=\{\psi_{\rm TF}, 0\}$ (as discussed above). It is well-known \cite{emergent,revnonlin} that the harmonic potential introduces a discrete (point) BdG spectrum for this spatially confined ground state. In addition to that, the translational invariance of the unconfined system is broken and, due to the presence of the DB soliton, a single eigenvalue $\lambda^{\rm (AM)}$ emerges. The respective (negative energy) eigenmode is the so-called anomalous mode (AM), while the associated eigenvalue $\lambda^{\rm (AM)}$ is directly connected with the oscillation frequency of the DB soliton in the harmonic trap, similarly to the case of a dark soliton in one-component BECs \cite{fms}. In fact, the imaginary part of the eigenvalue $\lambda^{\rm (AM)}$ reads $\lambda_i^{\rm (AM)} =\omega_{\rm osc}$, where $\omega_{\rm osc}$ is the oscillation frequency of the single DB soliton, given by \cite{BA}:
\begin{eqnarray}
\omega_{\rm osc}^2&=& \Omega^2 \left(\frac{1}{2} - \frac{\chi}{\chi_{\rm o}} \right),
\label{baomega} \\
\chi &\equiv& \frac{N_2}{\sqrt{\mu_1}}, \qquad
\chi_{\rm o} \equiv 8\sqrt{1+\left(\frac{\chi}{4}\right)^2}
\label{baomega2}
\end{eqnarray}
%

\begin{figure}
\includegraphics[width=5.6cm]{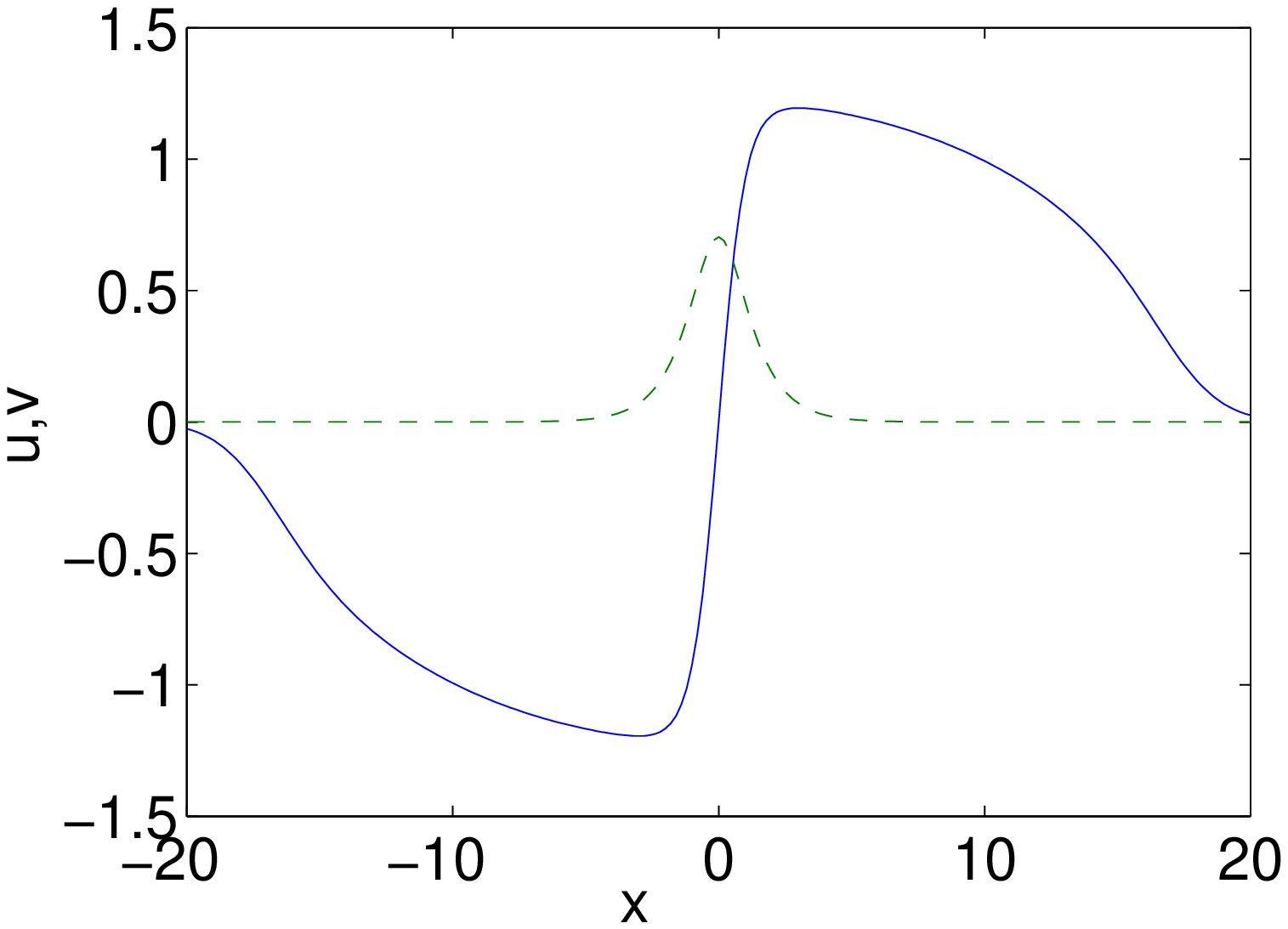}
\includegraphics[width=5.6cm]{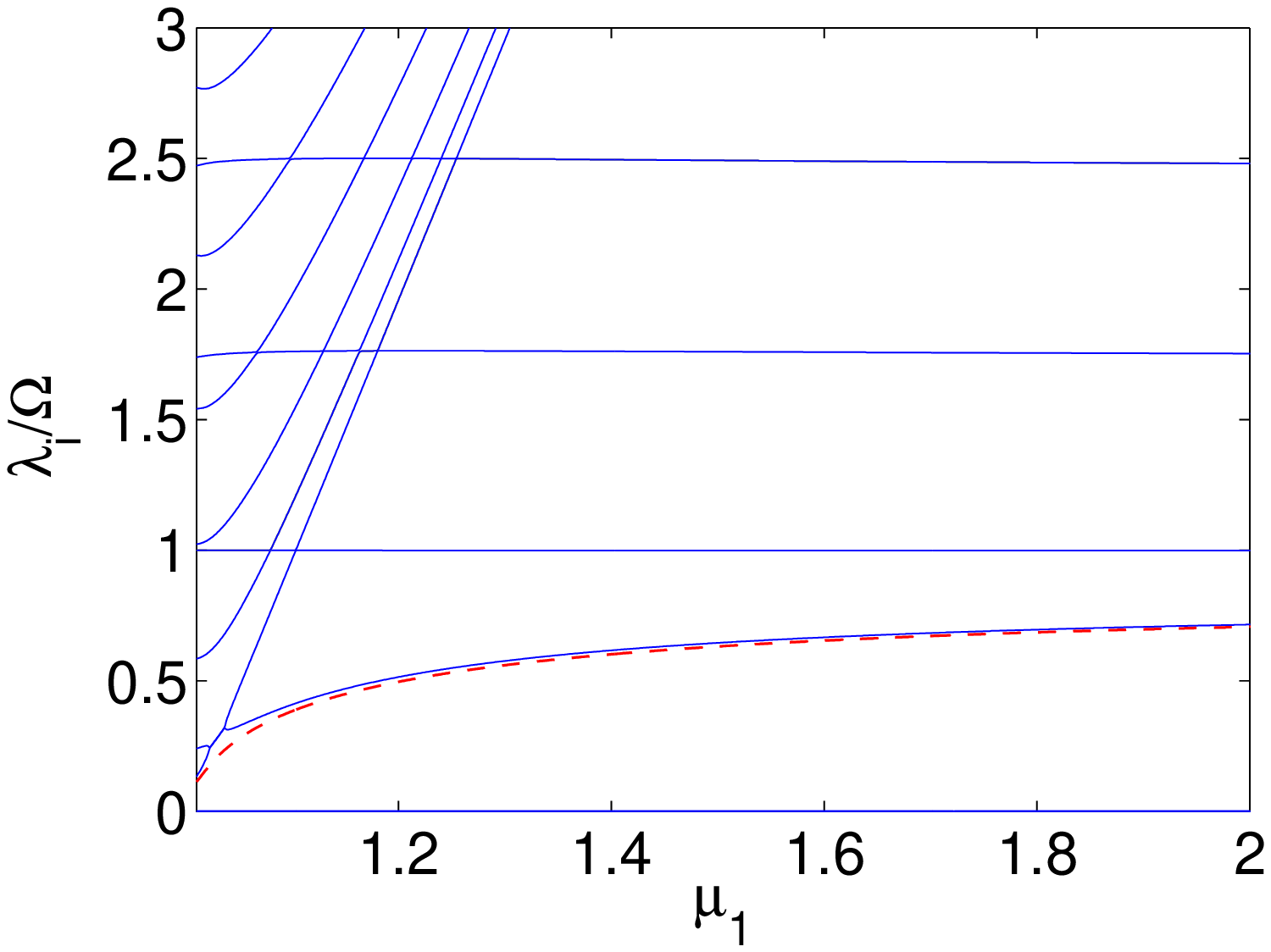}
\includegraphics[width=5.6cm]{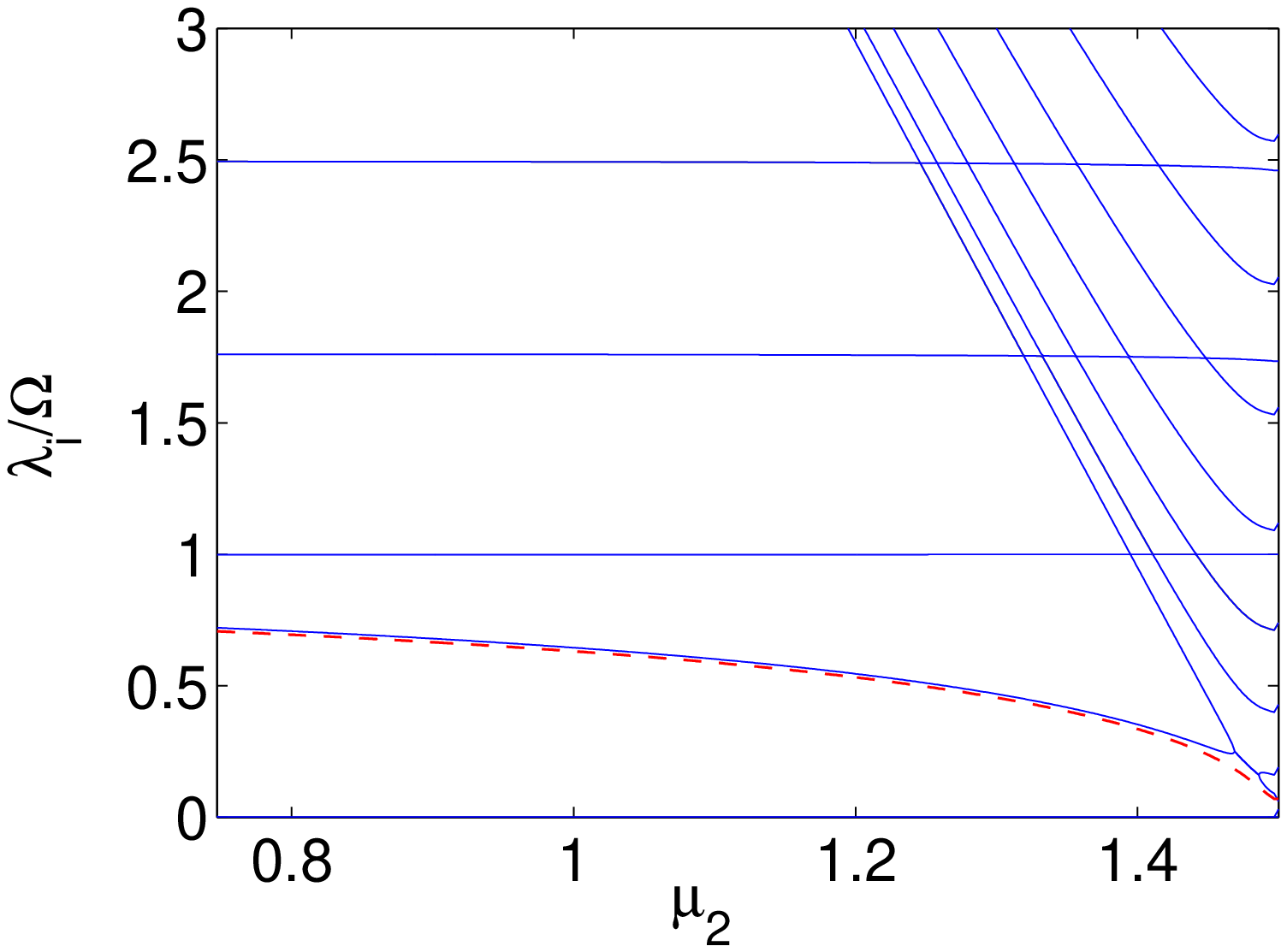}
\caption{(Color online)
The top panel depicts the stationary solution for a single DB-soliton
for $\mu_1=3/2$, $\mu_2=1$, and $\Omega=0.1$. The bright (dark) components
are shown
by the dashed green (solid blue) lines.
The middle (bottom) panel shows the normalized imaginary part
$\lambda_i/\Omega$ of the eigenvalues for the single DB-soliton
as a function of $\mu_1$ ($\mu_2$) for $\mu_2=1$ ($\mu_1=3/2$).
The (red) dashed line, depicts the analytical prediction of
Ref.~\cite{BA} for the DB-soliton oscillation frequency [cf. Eq.~(\ref{baomega})],
providing an excellent approximation to the anomalous mode eigenfrequency.}
\label{fig1}
\end{figure}

The above results are illustrated in Fig.~\ref{fig1}, where a typical example of a stationary single DB-soliton state is depicted (top panel); additionally, the eigenvalues $\lambda_i$ characterizing the excitation (BdG) spectra of such stationary states, are shown as functions of the chemical potentials $\mu_1$ and $\mu_2$ in the middle and bottom panels of the figure, respectively.
%
%
As observed in these two bottom panels, there exist two types of spectral lines, namely ``slowly-varying'' ones (analogous to ones that are present in the spectrum of a dark soliton in one-component BECs \cite{kip}) and ``fast-varying'' ones due to the
presence of the second (bright-soliton) component. The latter, as was 
pointed out also in the recent work of Ref.~\cite{engels3} may, in fact, 
collide with the internal anomalous mode of the DB soliton and give rise to instability
quartets which are barely discernible in Fig.~\ref{fig1} (see, e.g., the bottom panel for $\mu_2 > 1.4$ where a merger of eigenvalues occurs). Generally, however, it is found that the analytical prediction (red dashed line) is {\it excellent} in capturing the relevant anomalous mode pertaining to the DB-soliton oscillation.

The above discussion sets the stage for the presentation of our results for multiple DB-soliton states.

\section{Interaction between two dark-bright solitons}

We start by considering the case where the external trap is absent, i.e., for $\Omega=0$. To study the interaction of two identical DB solitons, as described in the ansatz of Eqs.~(\ref{eq15})-(\ref{eq16}), we will employ the Hamiltonian approach in the framework of the adiabatic approximation of the perturbation theory for matter-wave solitons (see, e.g., Refs.~\cite{revnonlin,djf}). In particular, we assume that the approximate two-DB-soliton state features an adiabatic evolution due to a weak mutual interaction between the constituent solitons and, thus, the DB soliton parameters become slowly-varying unknown functions of time $t$. Thus, $\phi \rightarrow \phi(t)$, $D \rightarrow D(t)$ and, as a result, Eqs.~(\ref{width})-(\ref{x0}) become:
\begin{eqnarray}
D^2(t)&=&\cos^2\phi(t)-\frac{1}{4}\chi D(t),
\label{s1} \\
\dot{x}_0(t)&=&D(t)\tan\phi(t),
\label{s2}
\end{eqnarray}
where we have used Eq.~(\ref{Nbmod}). The evolution of the parameters $\phi(t)$, $D(t)$ and $x_0(t)$ can then be found by means of the evolution of the DB soliton energy as follows. First, we substitute the ansatz (\ref{eq15})-(\ref{eq16}) into Eq.~(\ref{energy}) and perform the integrations under the assumption that the soliton velocity is sufficiently small, such that $\cos(kx) \approx 1$ (and $\sin(kx) \approx 0$). Then, we further simplify the result assuming that the solitons are  well-separated, i.e., their relative distance is $x_0\gg 1$. This way, we find that the total energy of the system assumes the following form:
\begin{eqnarray}
E = 2E_1+ E_{\rm DD}+ E_{\rm BB} + 2E_{\rm DB},
\label{energyf}
\end{eqnarray}
where $E_1$ is the energy of a single DB soliton, namely,
\begin{eqnarray}
E_1 =\frac{4}{3}D^3 + \eta^2 \left(\frac{k^2 - 2\left(\tilde{\mu}-1\right)}{D}+D \right),
%
\label{E1}
\end{eqnarray}
while the remaining terms account for the interaction between the two DB solitons. In particular, $E_{\rm DD}$, $E_{\rm BB}$, and $E_{\rm DB}$ denote, respectively, the interaction energy between the two dark solitons, the two bright ones, and the interaction energy between the dark soliton of one component and the bright one in the other component. The above interaction energies are given by the following (approximate) expressions:
\begin{eqnarray}
E_{\rm DD}&=& 16 \cos^2\phi \left[ \frac{1}{3}D \cos^2\phi +D +2(\cos^2\phi -D^2) x_0\right. \nonumber \\[1.0ex]
&&-\left. \frac{3+4\cos^2\phi}{3D} \cos^2\phi \right]{\rm e}^{-4Dx_0},
\label{edd}
\\[3.0ex]
E_{\rm BB}&=&\chi \Big[ 2D \left(D\left(1-Dx_0\right) -k^2 x_0\right) +D  \chi \Big]\nonumber \\
&\times&\cos\Delta\theta {\rm e}^{-2Dx_0} \nonumber \\
&+&\chi \Big[ \chi D\left(2Dx_0-1\right)\left(1+2\cos^2\Delta\theta\right) \Big]{\rm e}^{-4Dx_0},
\label{ebb}
\\[3.0ex]
E_{\rm DB}&=&-4\chi \cos^2\phi \cos\Delta\theta {\rm e}^{-2Dx_0} \nonumber \\
&+&\chi \cos^2\phi \Bigg[\frac{16}{3}\cos^2\phi-16Dx_0+8 \Bigg]{\rm e}^{-4Dx_0},
\end{eqnarray}
where terms of order $O(e^{-6Dx_0})$ and higher have been neglected (nevertheless, it has been checked that their contribution does not alter the main results that will be presented below).

Having determined the two-DB-soliton energy [up to order $O(e^{-6Dx_0})$], we can find the evolution of the soliton parameters from the energy conservation, $dE/dt=0$. In fact, we focus on the case of low-velocity, almost black solitons (with $\dot{D}(t) \approx 0$ and $\cos\phi(t) \approx 1$), for which energy conservation leads to
%
%
the following nonlinear evolution equation for the DB soliton center:
\begin{eqnarray}
\ddot{x}_0 &=& F_{\rm int},
\label{eqmot} \\[1.0ex]
F_{\rm int}& \equiv &F_{\rm DD}+F_{\rm BB}+2F_{\rm DB}.
\label{Fint}
\end{eqnarray}
In the above equations, $F_{\rm int}$ is the interaction force between the two DB solitons (depending on
the soliton coordinate $x_0$), which contains the following three 
distinct contributions: the interaction forces
$F_{\rm DD}$ and $F_{\rm BB}$ between the two dark and two 
bright solitons, respectively, as well as the
interaction force $F_{\rm DB}$ of the dark soliton of the
one soliton pair with the bright soliton of the other pair.
These forces have the following form:
\begin{eqnarray}
F_{\rm DD}&=&\frac{1}{\chi_{\rm o}}\left[\frac{1}{3}(544-352D_0^2) +128D_0\left( D_0^2-1 \right)x_0 \right] \nonumber \\
&\times& {\rm e}^{-4D_0x_0},
\label{fdd} \\[2.0ex]
F_{\rm BB}&=&\frac{\chi}{\chi_{\rm o}} \Big[ -6D_0+4D_0^2x_0-2\chi \Big] \nonumber\\
&\times&D_0^2\cos\Delta\theta{\rm e}^{-2D_0x_0} \nonumber\\
&+& \frac{\chi^2}{\chi_{\rm o}} \Big[ \left(1+2\cos^2\Delta\theta\right)\left(-8D_0x_0+6\right)\Big] \nonumber\\
&\times&D_0^2{\rm e}^{-4D_0x_0},
\label{fbb} \\[2.0ex]
F_{\rm DB}&=&\frac{\chi}{\chi_{\rm o}}\Big[ 8D_0\cos\Delta\theta \Big]{\rm e}^{-2D_0x_0} \nonumber\\
&+&\frac{\chi}{\chi_{\rm_o}}\Big[ -\frac{208}{3}+64D_0x_0 \Big]D_0{\rm e}^{-4D_0x_0},
\label{fdb}
\end{eqnarray}
where $D(t)\approx D_0$ since we are assuming that $\dot{D}(t)\approx 0$.

The equation of motion for the two-DB-soliton state [cf. Eq.~(\ref{eqmot})] 
provides a clear physical picture for the interaction between the two DB solitons. In order to better understand this result, first we note that the 
leading order interaction force between the bright soliton components 
is $\propto \exp(-2 D_0 x_0)$ and, as a result, it is decaying more slowly 
for large $x_0$ than the one
 between the two dark solitons 
which is $\propto \exp(-4 D_0 x_0)$; the interaction between dark and bright 
is also to leading order $\propto \exp(-2 D_0 x_0)$. 
This result is in accordance with earlier predictions, where the same 
dependence of the force over the soliton separation was 
found (see, e.g., Refs.~\cite{manton1} and \cite{darkint1,yuril,kip2} for 
bright and dark solitons, respectively).

Let us now consider the role of the bright-soliton component. In its absence, 
i.e., for $\chi=0$ [cf. Eq.~(\ref{s1})], it is clear that 
$F_{\rm BB}=F_{\rm DB}=0$ and Eq.~(\ref{eqmot}) describes the interaction 
between two dark (almost black) solitons; in this case, taking into regard 
that $D_0=1$, it can readily be found that the pertinent (repulsive) 
interaction potential is $\propto 2\exp(-4x_0)$, which coincides with the 
result of Ref.~\cite{yuril} obtained by means of a variational approach 
(see also a relevant discussion in Refs.~\cite{djf,kip2}). On the other hand, 
when bright solitons are present (i.e., for $\chi \ne 0$), the principal
nature of the 
bright-bright-soliton interaction ---and also of part of the 
dark-bright-soliton interaction one--- depends on the factor 
$\cos\Delta\theta$: when the relative phase between the bright-soliton 
components is $\Delta \theta=0$  ($\Delta\theta=\pi$), i.e., in the 
in-phase (out-of-phase) case, the interaction is repulsive (attractive). This 
conclusion stems from the fact that the coefficients of the terms 
$\propto \cos\Delta\theta$ are positive definite since the parameter 
$D_0 \ge 1$ for every $\chi>0$.

According to the above, it is clear that the competition between 
repulsive (for dark solitons) and attractive (for out-of-phase bright 
solitons) forces leads to the emergence of fixed points in the equation of 
motion (\ref{eqmot}) or, in other words, to the existence of 
a {\it stationary} two-DB-soliton state~\footnote{The presence of the
dark-bright interaction does not alter qualitatively (although it 
obviously affects quantitatively) this interplay between repulsive
terms $\propto \exp(-4 D_0 x_0)$ and attractive ones 
(if $\cos\Delta\theta=-1$) $\propto \exp(-2 D_0 x_0)$.}. Below we will 
demonstrate this effect in detail: we will determine the fixed (equilibrium) 
points, $x_{\rm eq}$, as solutions of the transcendental equation resulting 
from Eq.~(\ref{eqmot}) for $\ddot{x}_0=0$ in the out-of-phase case, and 
subsequently study their stability. Nevertheless, before proceeding further, 
we should mention that stationary two-DB-solitons were also found numerically 
and experimentally in Ref.~\cite{seg2} in the context of nonlinear optics, but 
their existence details and stability properties were not considered. 
Additionally, although exact two-DB-soliton solutions (as well as 
$\mathcal{N}$-DB-soliton solutions) do exist in the Manakov 
system \cite{rajendran,raj2}, their complicated form does not allow for a 
transparent physical description of the relevant dynamics, as provided above.

Let us now study the stability of the equilibrium points in the framework of Eq.~(\ref{eqmot}). Introducing the ansatz
$x_0(t) = x_{\rm eq} +\delta(t)$, and linearizing with respect to the small-amplitude perturbation $\delta(t)$, we derive the following equation:
\begin{eqnarray}
\ddot{\delta}+\omega_0^2\delta=0,
\label{eq27}
\end{eqnarray}
where the oscillation frequency $\omega_0$ is given by:
\begin{eqnarray}
\omega_0^2= - \frac{\partial F_{\rm int}}{\partial x_0} \bigg|_{x_0 = x_{\rm eq}},
%
\label{eq28}
\end{eqnarray}
%
where the phase-difference between the bright-soliton components is taken to be $\Delta \theta = \pi$.
Physically speaking, the oscillation frequency $\omega_0$ represents
the internal (out-of-phase) motion of the two DB-solitons; in fact, as here we deal with the
homogeneous case (i.e., in the absence of the trap), the in-phase motion
of the solitons is associated with the neutral translation mode
due to the translational invariance of the system (the respective in-phase
Goldstone mode has a vanishing
frequency).

\begin{figure}
\includegraphics[width=6.7cm]{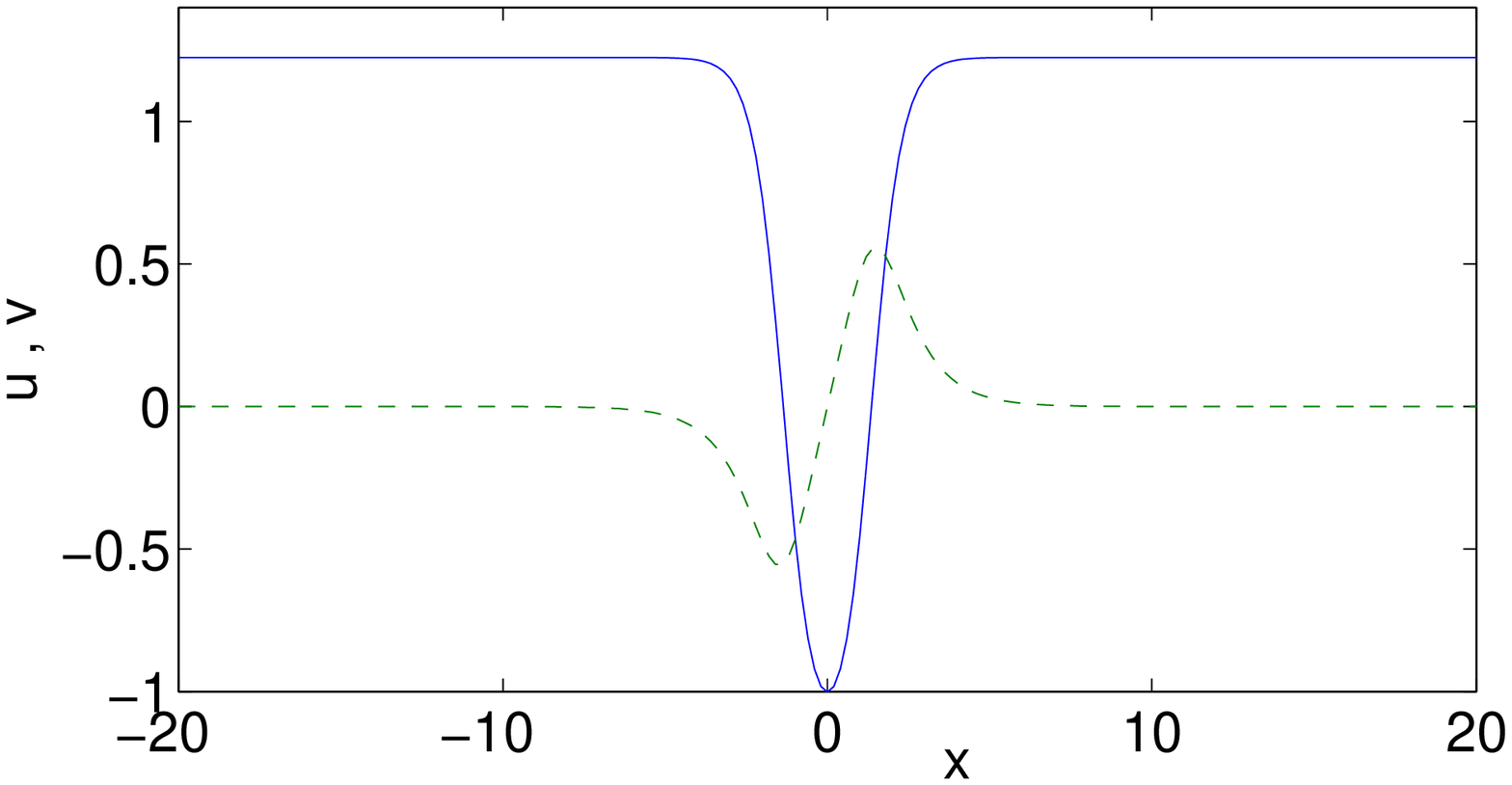}
\includegraphics[width=6.7cm]{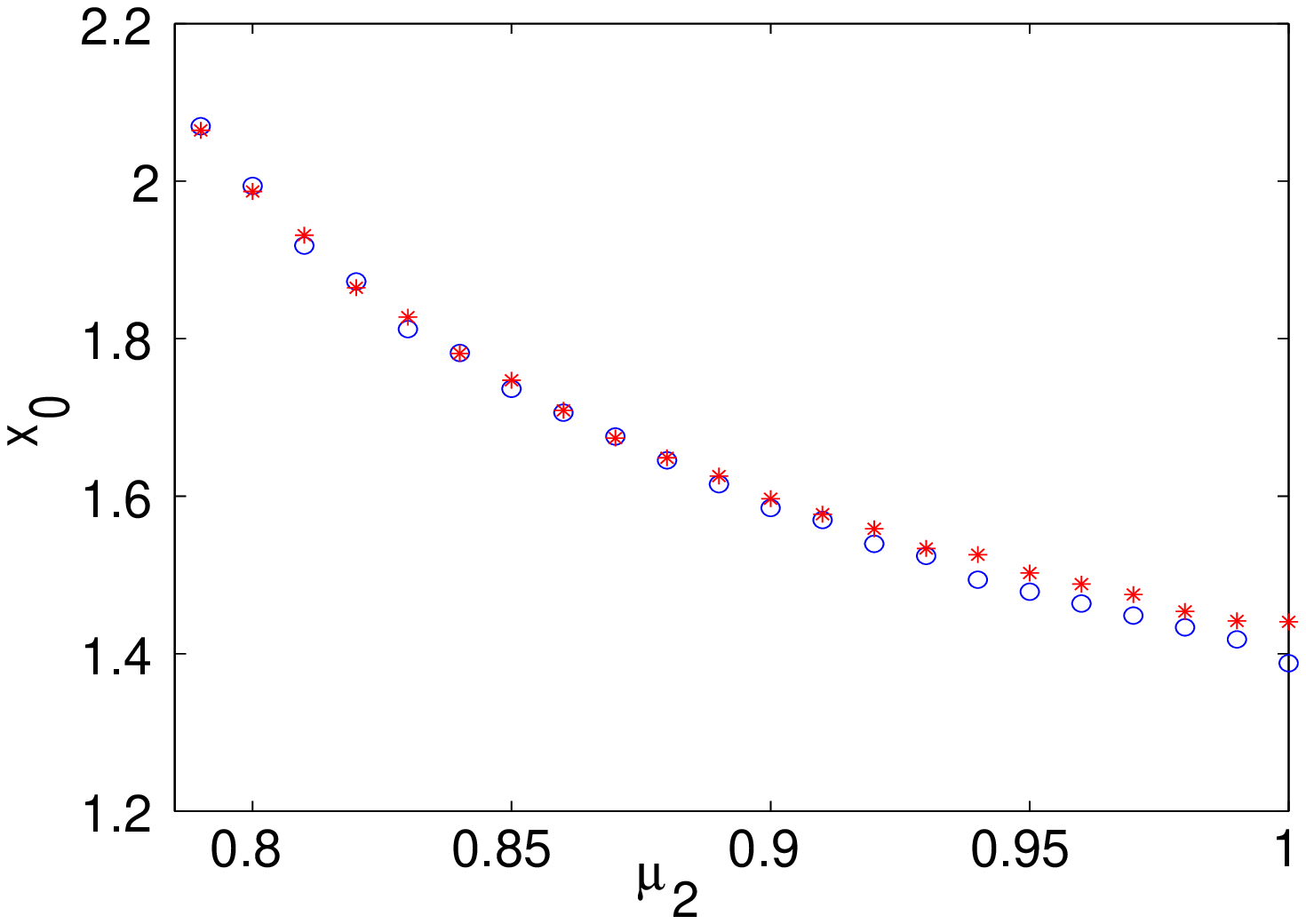}
\includegraphics[width=6.7cm]{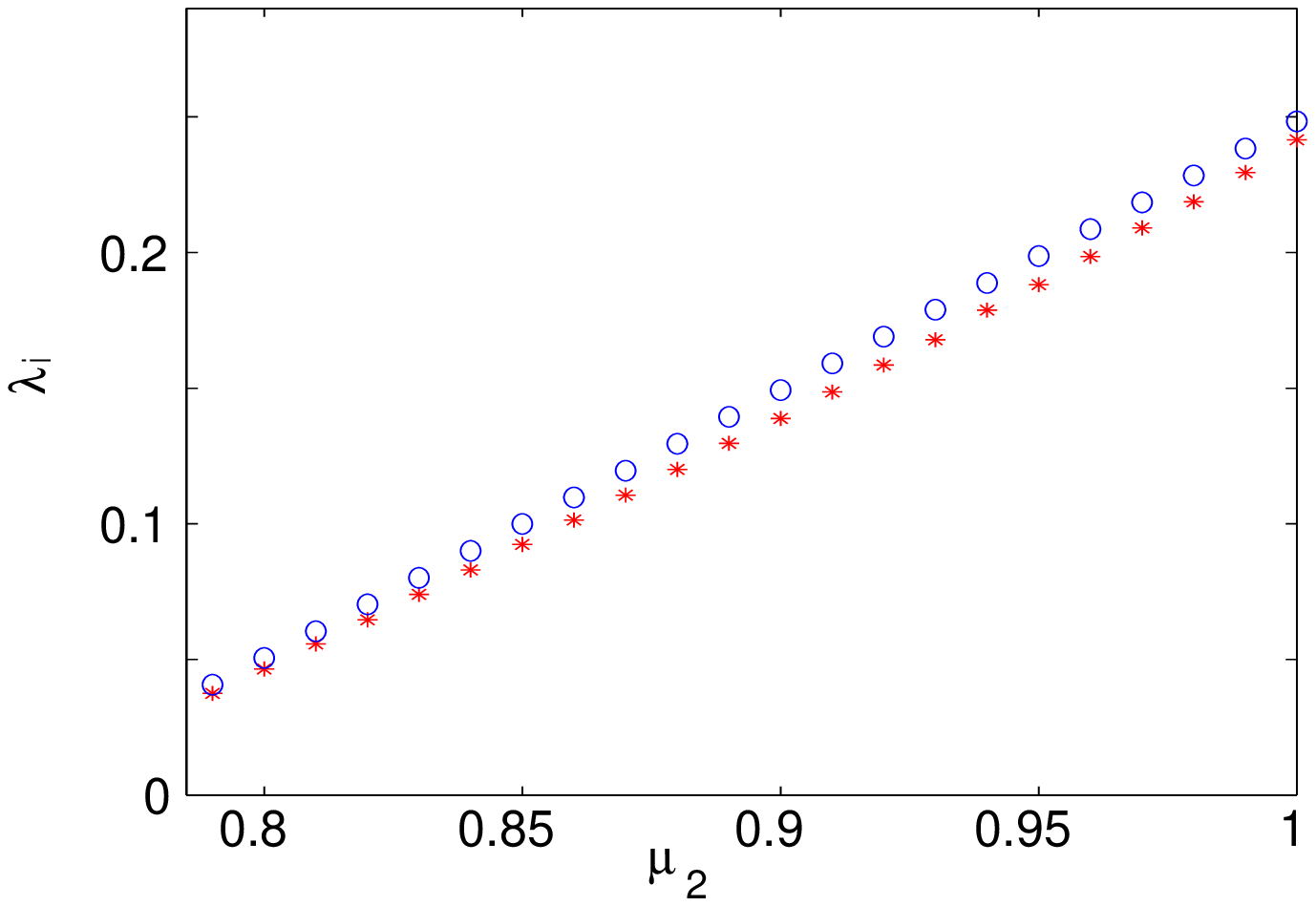}
\caption{(Color online)
Top panel: A stationary DB-soliton pair: the solid (blue) line denotes
the two-dark-soliton state [recall that each dark soliton is associated
with a zero crossing], while the dashed (green) line denotes the
respective two-bright-soliton state. The chemical potentials are
$\mu_1=3/2$ and $\mu_2=1$.
Middle panel: the equilibrium center of mass $x_0$ as a function of the chemical
potential $\mu_2$ (for $\mu_1=3/2$). Stars (in red) denote the analytical
prediction of Eq.~(\ref{eqmot}), while circles (in blue) denote the
numerically obtained soliton center $x_0$.
Bottom panel: the oscillation frequency for the out-of-phase motion of
the DB-soliton pair as a function of $\mu_2$ (for $\mu_1=3/2$).
Stars (in red) depict the analytical result for
$\omega_0$ [cf. Eq.~(\ref{eq28})], while circles (in blue) depict
the numerically obtained imaginary eigenvalue $\lambda_i$
(for the out-of-phase soliton motion) of the excitation spectrum.}
\label{fig3}
\end{figure}

The above analytical predictions have been compared with numerical simulations. First, we have confirmed the existence of the stationary two-DB-soliton state (in the out-of-phase case); a prototypical example of such a state is shown in the top panel of Fig.~\ref{fig3} (for $\mu_1=3 \mu_2/2=3/2$). We have also determined the dependence of the equilibrium soliton positions 
(denoted by $x_0$ in the middle panel of Fig.~\ref{fig3}) and the effective 
frequency $\omega_0$ [cf. Eq.~(\ref{eq28})] on the chemical potential $\mu_2$ 
of the bright soliton component. The respective analytical and numerical 
results are shown in the middle and bottom panels of Fig.~\ref{fig3}.
To obtain the numerical results, we have used a (least squares) fitting algorithm to accurately identify the amplitude $\eta$, inverse width $D$, and equilibrium center of mass $x_0$ of the bright component.
The numerical findings for $x_0$ and $\omega_0$ (the latter is obtained via a 
BdG analysis, as the imaginary eigenvalue $\lambda_i$ of the stationary 
two-DB-soliton state) are directly compared with the semi-analytical results 
of Eqs.~(\ref{eqmot}) and (\ref{eq28}), respectively. Taking into account the 
approximate nature of the fitting scheme, we find that there is a very good 
quantitative agreement between the analytical and numerical results 
(see middle and bottom panels of Fig.~\ref{fig3}).
Notice that despite the motion of this eigenvalue through the continuous 
spectrum,
no instability is observed in the parametric window shown in Fig.~\ref{fig3}.

\section{Multiple dark-bright solitons in the trap}

Next, let us consider the case of multiple DB-solitons in the presence of the harmonic trap. In the presence of the trap, each of the multiple-DB-soliton structures is subject to two forces: (a) the restoring force of the trap, $F_{\rm tr}$ [in the case of a single DB-soliton, this force induces an in-trap oscillation with a frequency $\omega_{\rm osc}$ ---see Eq.~(\ref{baomega})], and (b) the pairwise interaction force $F_{\rm int}$ [cf.~Eq.~(\ref{Fint})] with other dark-bright solitons. Thus, taking into regard that $F_{\rm tr}=-\omega_{\rm osc}^2 x_0$ \cite{BA}, one may write the effective equation of motion for the center $x_0$ of a two-DB-soliton state as follows:
\begin{equation}
\ddot{x}_0 = F_{\rm tr} + F_{\rm int}.
\label{eqmot2}
\end{equation}
One can thus straightforwardly generalize the above equation for $\mathcal{N}$-interacting DB-soliton states, similarly to the case of multiple dark solitons in one-component BECs \cite{kip,kip2,coles}.

It is interesting to observe that, in the presence of the trap, the restoring force $F_{\rm tr}$ can generate equilibrium positions, not only for out-of-phase bright solitons (whereby such a state could be stationary even without the trap as found above), but also for in-phase bright solitons. In the latter case, the repulsion between both  the dark- and the bright-soliton component(s), is 
balanced by the trap-induced restoring force. In the case of two-DB solitons placed at $x=\pm x_0$, the equilibrium points, $x_{\rm eq}$, can readily be found (as before) as solutions of the transcendental equation resulting from Eq.~(\ref{eqmot2}) for $\ddot{x}_0=0$, in both the in- and out-of-phase cases. To study the stability of these equilibrium points in the framework of Eq.~(\ref{eqmot2}), we may again use the ansatz $x_0(t) = x_{\rm eq} +\delta(t)$, and obtain a linear equation for the small-amplitude perturbation $\delta(t)$, similar to that of Eq.~(\ref{eq27}), namely:
$\ddot{\delta}+\omega_1^2 \delta=0$, where the frequency $\omega_1$ is given by,
\begin{eqnarray}
\omega_1 ^2 &=& \omega_{\rm osc}^2 + \omega_0^2,
\label{eq29}
\end{eqnarray}
where $\omega_0$ is given by Eq.~(\ref{eq28}). Similarly to the case of dark solitons in one-component BECs \cite{kip2} (see also Ref.~\cite{djf}), by construction, this mode captures the out-of-phase motion of the DB-soliton pair. Furthermore, by symmetry, the in-phase oscillation of the DB-soliton pair in the trap will be performed with the frequency $\omega_{\rm osc}$. These two characteristic frequencies coincide with the eigenfrequencies of two anomalous modes of the 
BdG spectrum of the trapped DB-soliton pair
(see, e.g., Refs.~\cite{djf,kip2,law} for a relevant discussion of such modes). The rest of the spectrum will still be discrete (due to the presence of the 
harmonic trap ---see also Sec.~III), but in this case also collisions of these 
anomalous modes of the DB-solitons with the modes of the background may 
induce instabilities (see also Ref.~\cite{kip2}) ---see below.

\begin{figure}
\includegraphics[width=7cm]{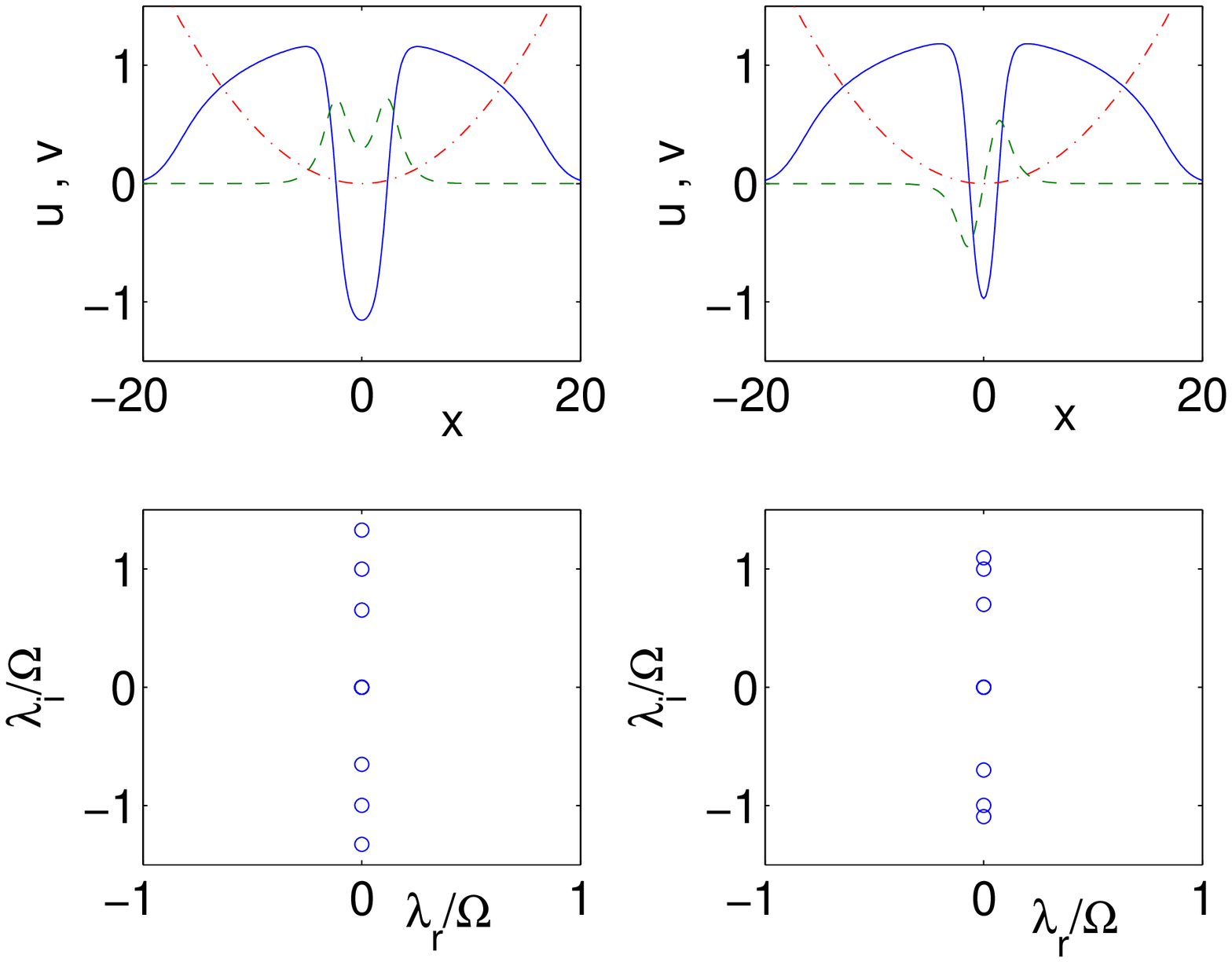}
\includegraphics[width=3.5cm]{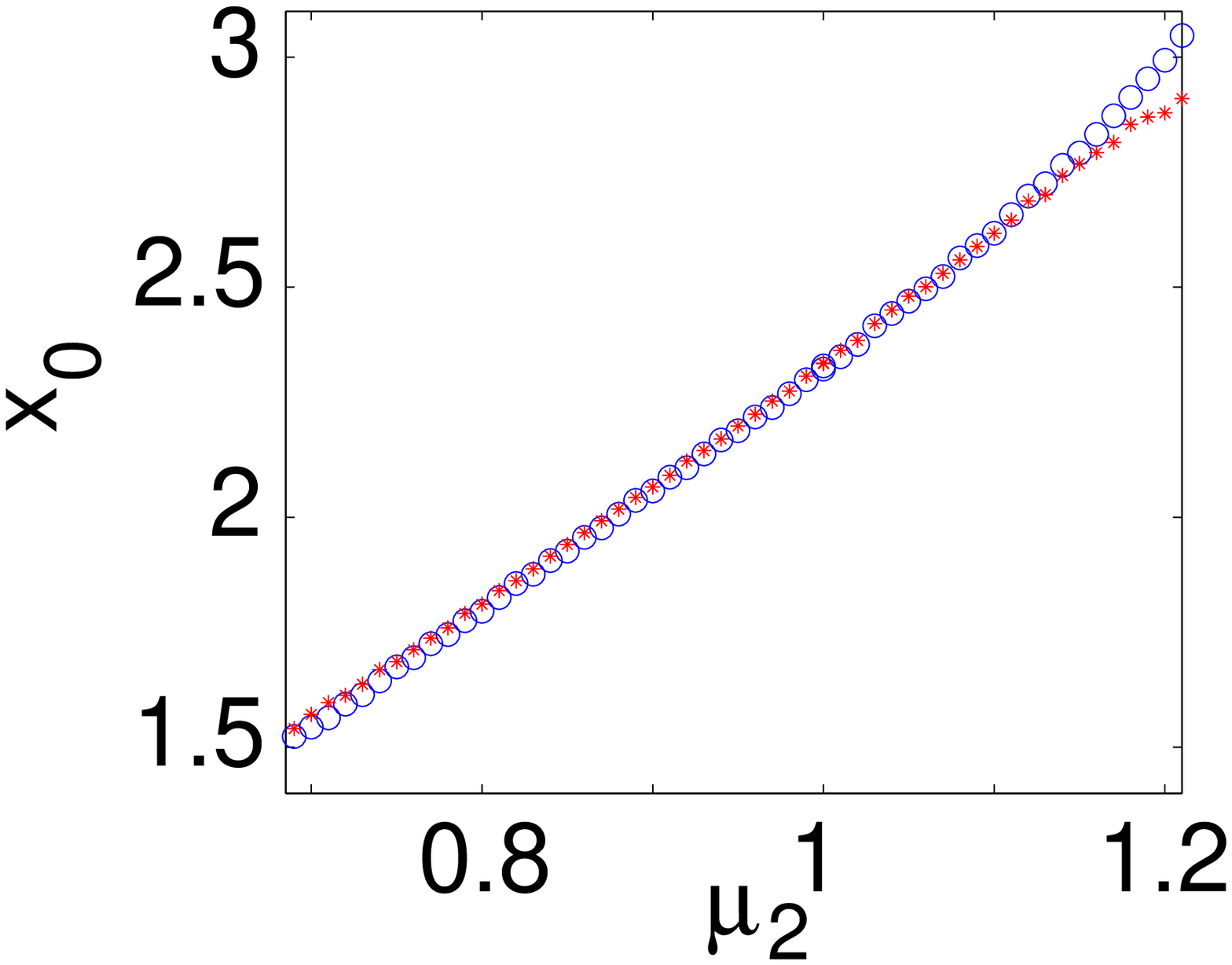}
\includegraphics[width=3.5cm]{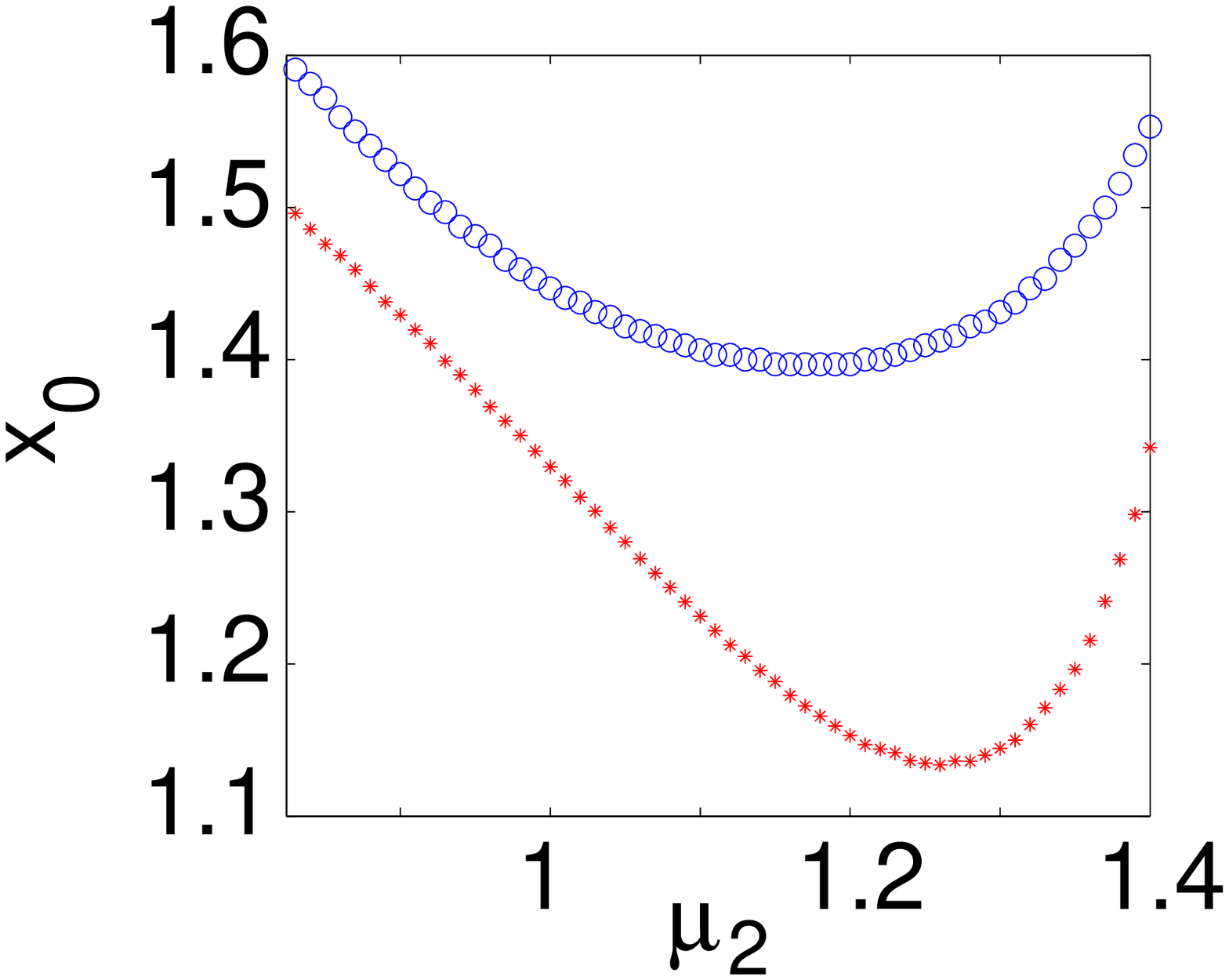}
\caption{(Color online)
The left and right columns correspond, respectively, to an
in-phase and an out-of-phase dark-bright soliton pair in
a harmonic trap with $\Omega=0.1$.
The top row of panels depicts the profiles of the DB-soliton pairs
(solid blue lines and dashed green lines corresponding, respectively,
to the dark and bright components) and the trapping potential (dashed-dotted red line).
The middle row of panels depicts
the spectral plane $(\lambda_r,\lambda_i)$ rescaled by the trap frequency $\Omega$.
The bottom row of panels depicts the numerical (small stars in red) and
the analytical (circles in blue) results for the equilibrium distance between
the solitons as a function of $\mu_2$;
the theoretical prediction is based on Eq.~(\ref{eqmot2}).}
\label{fig4}
\end{figure}

\begin{figure}
\begin{center}
\includegraphics[width=4.1cm]{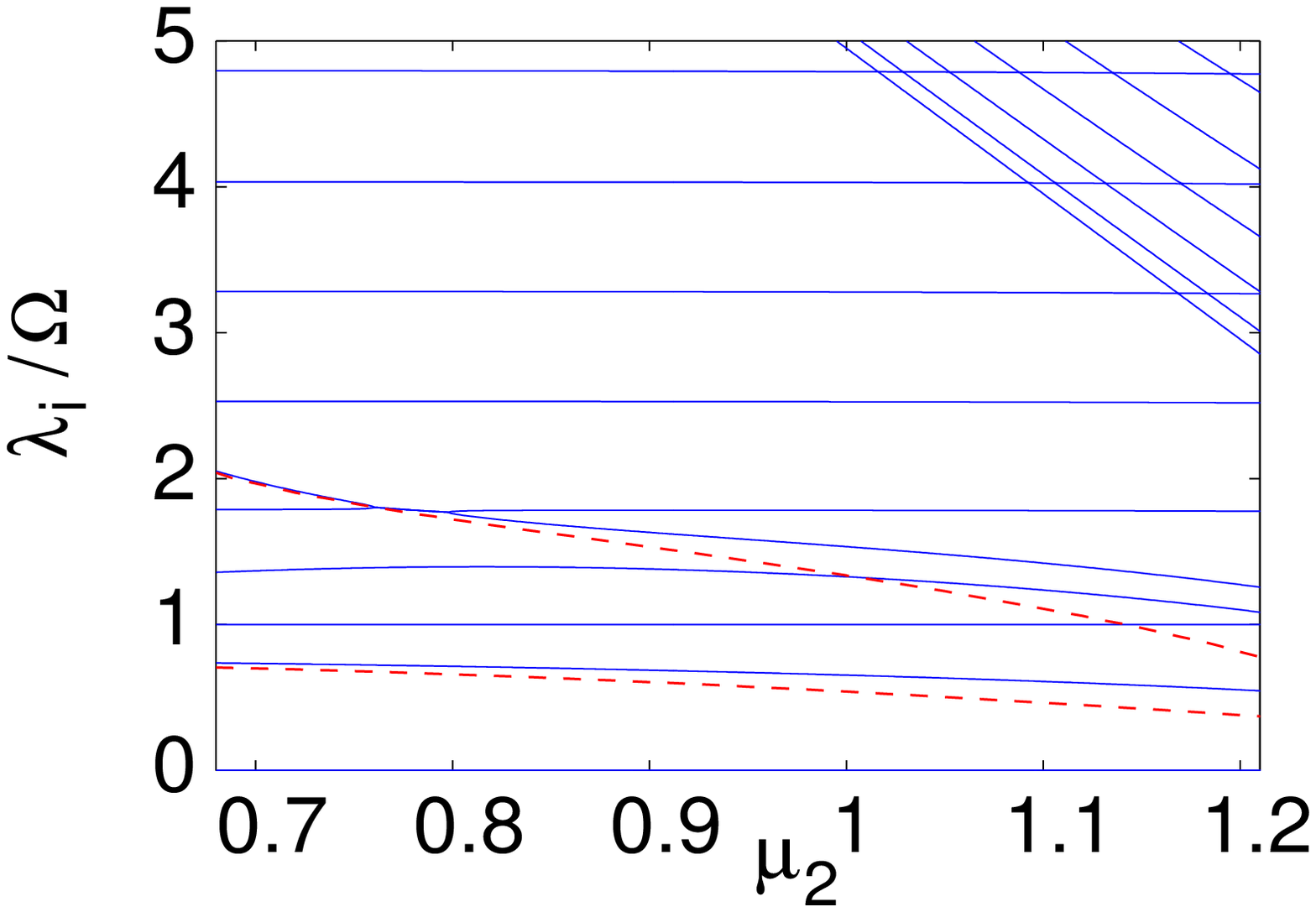}
\includegraphics[width=4.1cm]{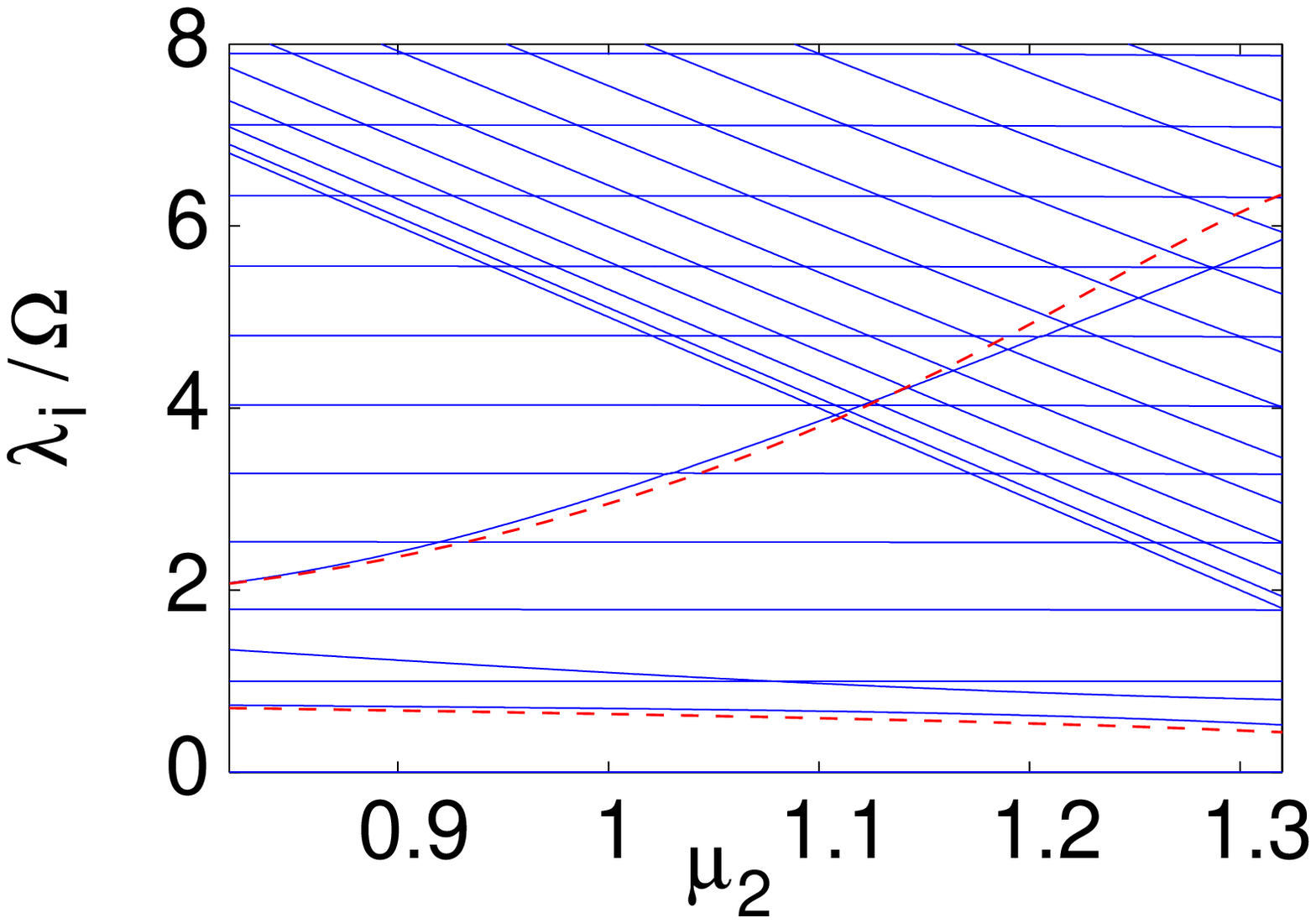}
\\
\includegraphics[width=8.45cm,height=3.25cm]{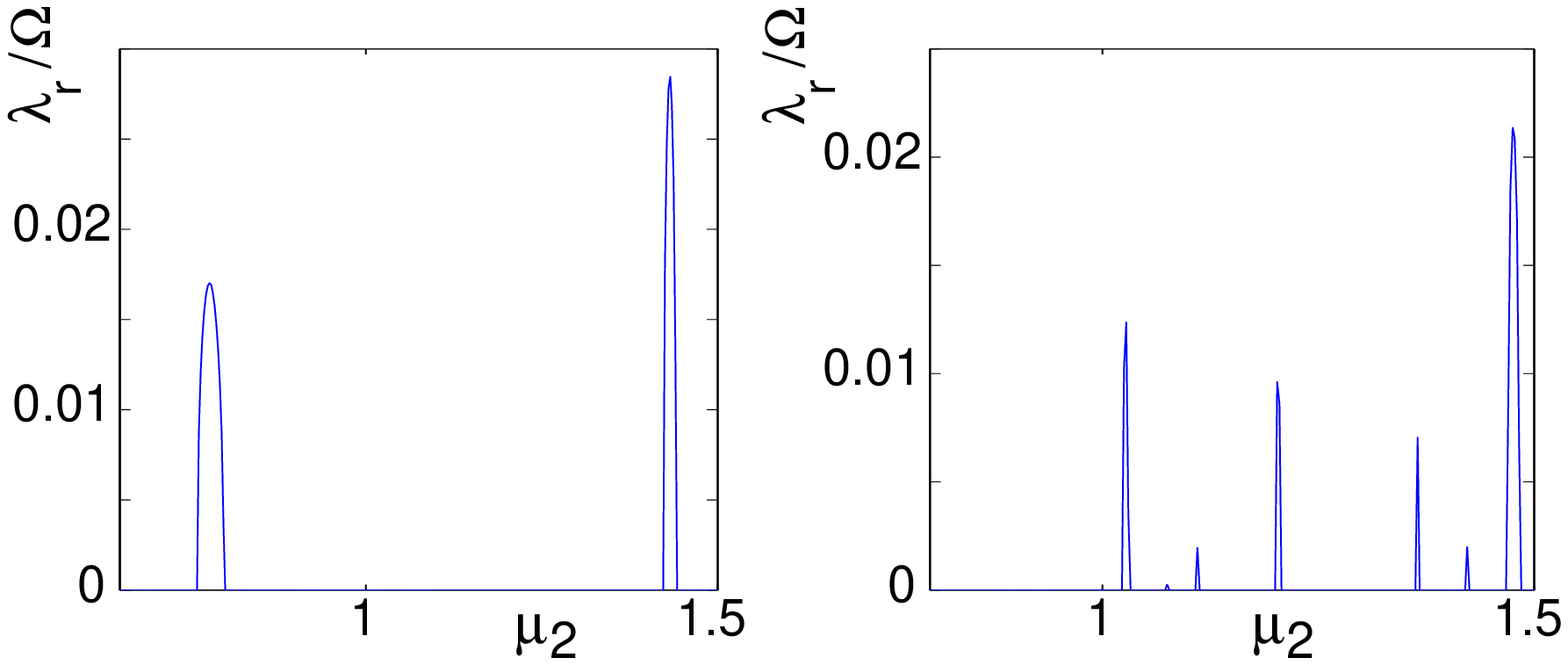}
\end{center}
\caption{(Color online)
The left and right columns of panels correspond, respectively, to an in-phase
and an out-of-phase dark-bright soliton pair in a harmonic trap with $\Omega=0.1$.
Shown are the imaginary (top row of panels) and the real (bottom row of panels)
parts of the eigenvalues as functions of $\mu_2$ for $\mu_1=3/2$.
In the top panels, the theoretical predictions for the eigenfrequencies of
the anomalous modes of the system, pertaining to the in- and out-of-phase
oscillations of the DB-solitons are depicted by dashed (red) lines.
Notice that collisions of modes (eigenvalue crossings) observed in the
top panels indicate the emergence of instability windows observed in the
bottom panels. The instabilities are of the Hamiltonian-Hopf type and
result in the emergence of eigenvalue quartets.}
\label{fig5}
\end{figure}

\begin{figure}
\includegraphics[width=8.5cm]{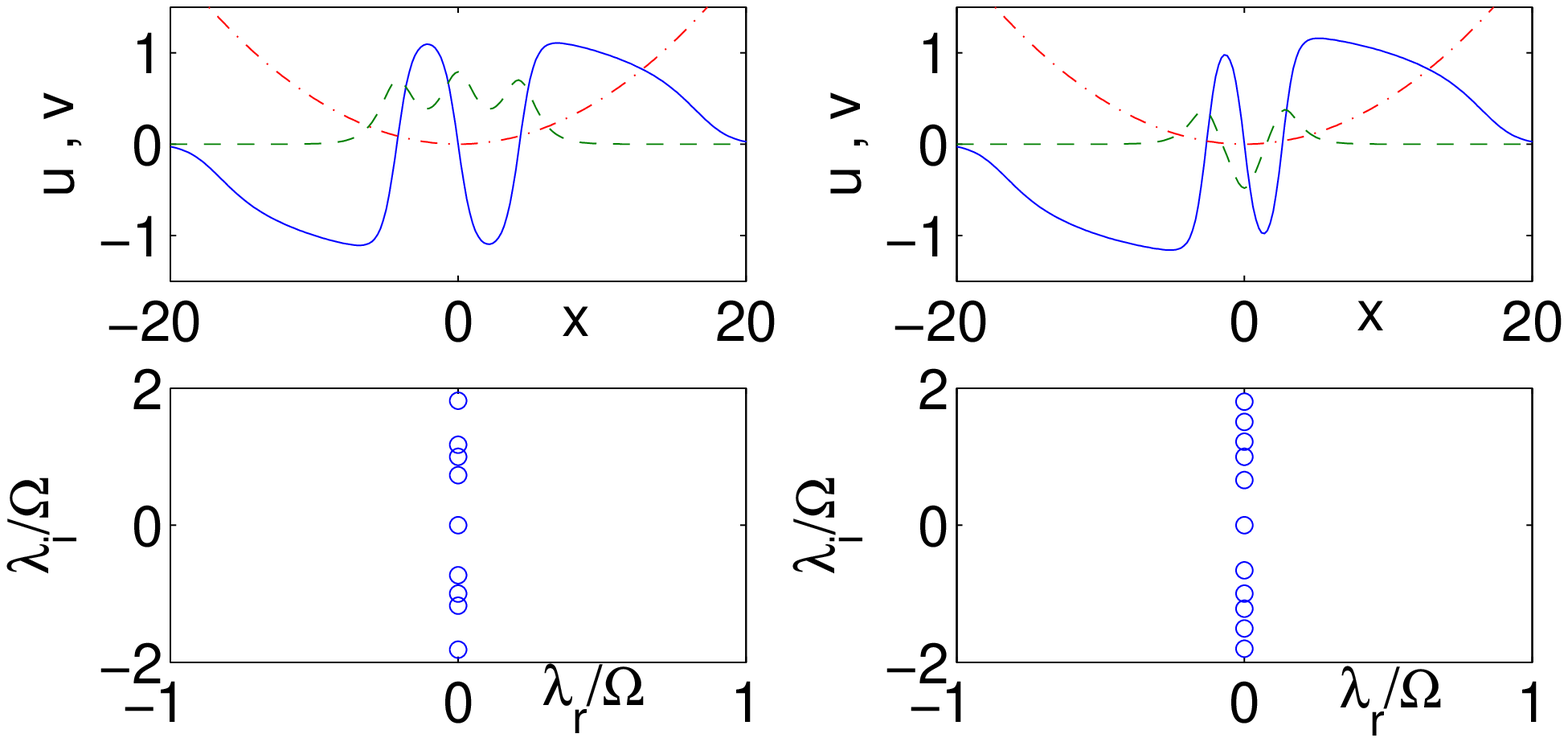}
\includegraphics[width=8.5cm,height=5.35cm]{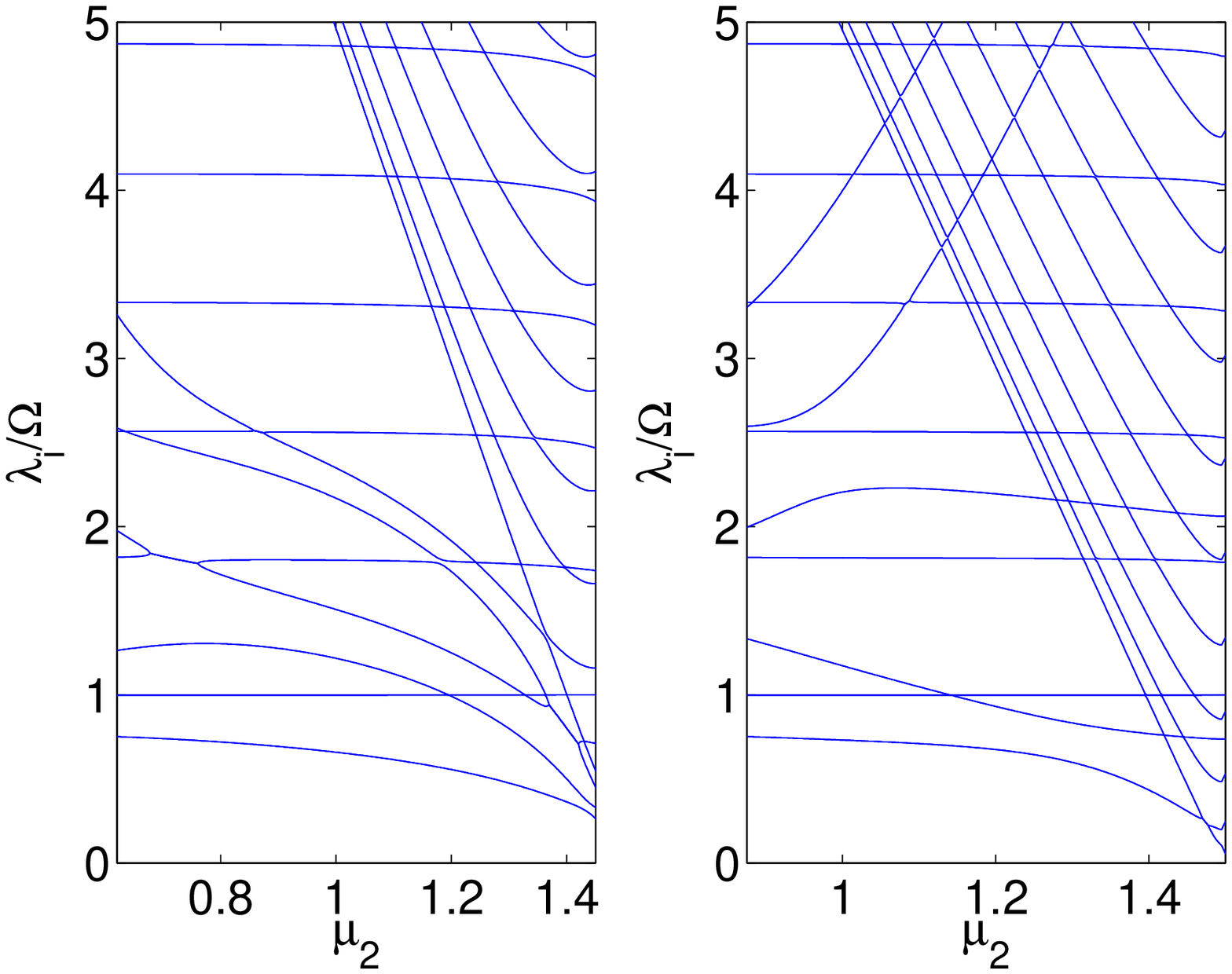}
\includegraphics[width=8.5cm,height=5.35cm]{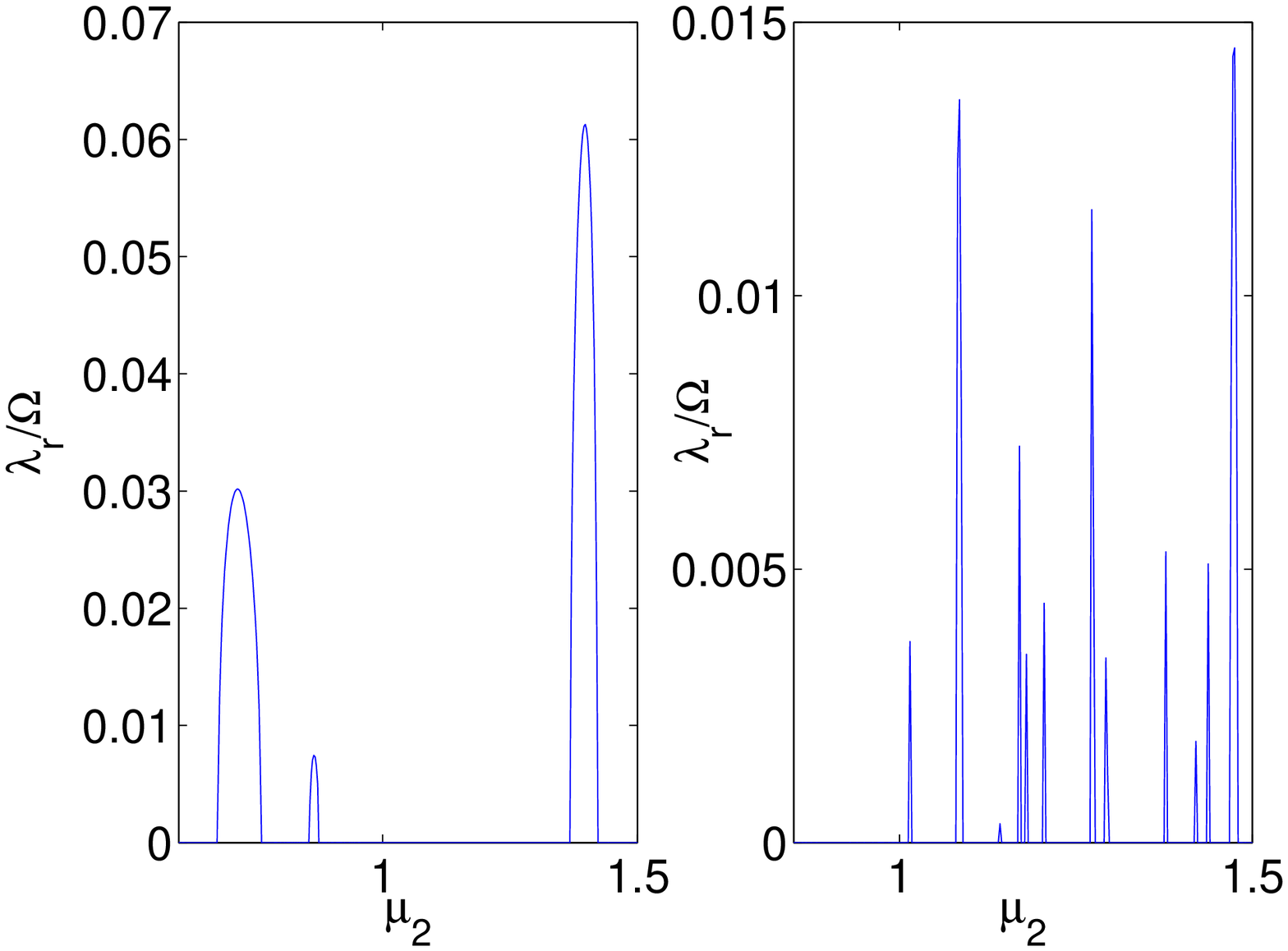}
\caption{(Color online)
The left and right columns of panels correspond, respectively, to an in-phase and
an out-of-phase three-DB-soliton configurations.
The top row of panels depicts the respective stationary states,
for $\mu_1=3/2$, $\mu_2=1$ and $\Omega=0.1$; solid (blue) lines depict the
dark-soliton components, dashed (green) lines the bright ones,
while the dashed-dotted (red) line shows the harmonic trap.
The second row of panels depicts the spectral planes for the
above stationary states, while the third and fourth rows of panels are
equivalent to those of Fig.~\ref{fig5},
but for the three-DB-soliton configurations.}
\label{fig7}
\end{figure}

We now turn to a systematic numerical investigation of the above features and of the multiple-DB-soliton states. At first, we consider the two-DB-soliton state in the trap, results for which are summarized in Figs.~\ref{fig4} and \ref{fig5}, both for the in-phase and the out-of-phase configurations. In particular, the top left and right panels of Fig.~\ref{fig4} show examples of an in-phase and an out-of-phase stationary DB-soliton pair, respectively (both for $\mu_1=3/2$ and $\mu_2=1$). The two middle panels illustrate the corresponding spectral planes, showcasing the linear stability of these configurations. The bottom panels of the figure show the equilibrium positions of the soliton centers. In the in-phase case (bottom left panel), it is observed that larger chemical potential (number of atoms) in the second component leads to stronger repulsion and, hence, larger distance from the trap center. In the out-of-phase case (bottom right panel), we observe a similar effect but in the reverse direction (due to the attraction of the out-of-phase bright-soliton components) for smaller values of the chemical potential. Notice that in both cases a good agreement is observed between the numerically observed equilibrium separations and the theoretically predicted ones from Eq.~(\ref{eqmot2}).

To study the validity of Eq.~(\ref{eq29}) ---pertinent to small-amplitude oscillations around the fixed points--- we show in Fig.~\ref{fig5} the eigenvalues $\lambda$ of the excitation spectrum [both for the in-phase (left column) and 
for the out-of-phase (right column) cases] as functions of $\mu_2$. The imaginary and real part, $\lambda_i$ and $\lambda_r$, of the respective eigenvalues, normalized over the trap strength $\Omega$, are respectively shown in the top and bottom panels of Fig.~\ref{fig5}. In the top panels, it is straightforward to compare the analytical result of Eq.~(\ref{eq29}) with the BdG result, namely the second anomalous mode of the spectrum, corresponding to the out-of-phase oscillations of the DB-soliton pair. Once again, good agreement is observed between the two; the differences may be partially attributed to the ``interaction'' (i.e., collisions) of these modes with other modes of the BdG spectrum. It is clear from the comparison of the corresponding columns that there exist narrow instability windows, arising due to the crossing of the anomalous mode(s) of the DB-soliton pair with eigenmodes of the background of the two-component system. These instabilities arise in the form of Hamiltonian-Hopf bifurcations \cite{HH} through the emergence of quartets of complex eigenvalues resulting from the collision of two pairs.
The growth rates of the pertinent oscillatory instabilities are fairly small (i.e., the instabilities are weak) in both the in- and out-of-phase cases; it should be noted, however, that in the latter case, the formation of the quartets appears to be occurring in very narrow intervals.

Naturally, the above considerations can also be generalized to three- or more DB-solitons, although the analytical calculations become increasingly more tedious; again, as we will show below, in-phase or out-of-phase configurations are possible in the presence of the trap. Pertinent examples, showing two different three-DB-soliton configurations, are illustrated in Fig.~\ref{fig7}. In particular, the first column in the figure corresponds to the in-phase three-DB-soliton state, while
the second column corresponds to the out-of-phase variant thereof. In the case under consideration, there exist narrow parametric intervals of dynamical instability, which are narrower for the out-of-phase case (as in the case of the two-DB-soliton states). We should mention, in passing, that the dynamics of two- and three-DB soliton configurations was recently studied in Ref.~\cite{berloff}; 
our study complements the latter by yielding analytical approximations 
and a numerical continuation/bifurcation approach towards such states.

\section{Conclusions and discussion}

In the present work, we have studied multiple quasi-one-dimensional dark-bright (DB) solitons in atomic Bose-Einstein condensates. Our theoretical results were motivated and supported by the experimental evidence of the formation of DB-soliton clusters in a two-component, elongated rubidium condensate, confined in a harmonic trap. The theoretical analysis was based on the study of two coupled, one-dimensional Gross-Pitaevskii equations.

Starting from the case of a homogeneous condensate (i.e., in the absence of a trapping potential), we have employed a Hamiltonian perturbation theory to analyze the interaction between two DB-solitons. Assuming that the DB-solitons are of low velocity and sufficiently far from each other, we have found approximate expressions for the interaction forces between the same or different soliton components. This way, we derived a classical equation of motion for the center of mass of the DB-soliton pair, and revealed the role of the phase-difference between the bright-soliton components: we have shown, in particular, that the repulsion between the dark soliton components may be counter-balanced by the attraction between out-of-phase bright components, thus inducing the existence of stationary DB-soliton pairs even in the case when the external trapping potential is absent. We have found the equilibrium distance between the two DB solitons that compose the stationary DB-soliton pair, with the semi-analytical result being in excellent agreement with the relevant numerical one. Additionally, we have demonstrated the linear stability of these stationary DB-soliton pairs by means of analytical and numerical techniques [the latter were based on a Bogoliubov-de Gennes analysis]. It was shown that the analytical result for the oscillation frequency of small-amplitude perturbations around the equilibrium distance is in excellent agreement with the pertinent eigenvalue characterizing the excitation spectrum of the DB-soliton pair.

We have then studied multiple-DB-solitons in the trap. In this case, we have 
employed a simple physical picture, where the total force acting on the 
DB-solitons was decomposed to an interaction force (derived in the 
homogeneous case) and a restoring force induced by the trapping potential; the 
relevant characteristic frequency associated with the latter was the 
oscillation frequency of a single-DB-soliton in the trap (which was found to 
coincide with the pertinent anomalous-mode eigenvalue of the single 
DB soliton system). Following this approach, we were able to find stationary 
in-trap DB-soliton pairs even in the case where the bright-soliton components 
were repelling each other: in this case, the trap-induced restoring force was 
able to counter-balance the repulsive forces between the dark- and the 
bright-soliton components. The semi-analytical results for the equilibrium 
distance and the oscillation frequencies (for the in- and out-of-phase cases) 
were again found to be in very good agreement with respective numerical 
results, including the anomalous modes' eigenfrequencies pertaining to the 
in- and out-of-phase motion of solitons. The stability analysis of the 
DB-solitons in the trap indicated the possibility of the existence of unstable 
modes through Hamiltonian-Hopf instability quartets, although the latter 
would typically only arise over narrow parametric intervals ---and with 
rather weak instability growth rates. Results pertaining to 
three-DB-solitons in the trap were presented as well; the main features of 
these states were found to be qualitatively similar to the ones of 
the DB-soliton pairs.

Coming back to our experimental findings, we should note the following. 
Given the nature of the available initial conditions,
our experimental observations were not able to identify, in a 
straightforward way, 
genuinely stationary DB-soliton complexes and/or to identify precisely their
internal modes. Nevertheless, the frequent and persistent occurrence of 
DB-soliton clusters in the experiment is highly indicative of the robustness 
of such ``DB-soliton molecules''. This is in tune with our existence and 
stability results.

It would be particularly interesting to further explore the dynamics of multiple-DB-soliton complexes, and potentially the formation of ``DB-soliton gases'' comprising such interacting atomic constituents. Deriving Toda-lattice-type equations describing such gases, and identifying their stationary states, excitations and (mesoscopic) solitons (as in the case of  single-component dark solitons \cite{coles}), would be challenges for future work. Another possibility is to extend the present considerations to the vortex-bright solitons found in Ref.~\cite{VB}. There, it would be relevant to identify whether molecular states consisting of two- or of three-vortex-bright solitons can be constructed, and whether the relative phases of $0$ and $\pi$ between the bright components can still yield different stationary states.
Relevant studies are presently in progress.

\section*{Acknowledgments}
P.G.K.\ acknowledges the support from NSF-DMS-0806762 and from the
Alexander von Humboldt Foundation.
The work of D.J.F. was partially supported by the Special Account for
Research Grants of the University of Athens.
R.C.G.\ gratefully acknowledges the support from NSF-DMS-0806762.
P.E. acknowledges
support from NSF and ARO.


\end{document}